\documentclass[a4paper, usenatbib]{mnras}
\usepackage{graphicx}
\usepackage{xcolor}
\usepackage{amsmath}
\usepackage{subfig}
\makeatletter
 \def\@textbottom{\vskip \z@ \@plus 1pt}
 \let\@texttop\relax
\makeatother

\title[Recurrence Analysis of AGN Variability]{Complex Variability of \textit{Kepler} AGN Revealed by Recurrence Analysis}

\author[R. A. Phillipson et al.]
  {R.\,A.~Phillipson\href{https://orcid.org/0000-0001-6891-7091}{\textcolor[HTML]{A6CE39}{\includegraphics[scale=0.5]{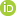}}},$^1$\thanks{E-mail: rebecca.a.phillipson@drexel.edu}
  P.\,T.~Boyd\href{https://orcid.org/0000-0003-0442-4284}{\textcolor[HTML]{A6CE39}{\includegraphics[scale=0.5]{ORCID_icon.png}}},$^2$
  A.\,P.~Smale,$^2$ M.\,S.~Vogeley\href{https://orcid.org/0000-0001-7416-9800}{\textcolor[HTML]{A6CE39}{\includegraphics[scale=0.5]{ORCID_icon.png}}}$^1$\\
  $^1$Department of Physics, Drexel University, 3141 Chestnut St, Philadelphia, PA 19104, USA\\
  $^2$Astrophysics Science Division, NASA Goddard Space Flight Center, Greenbelt, MD 20771, USA}
\date{Submitted to Monthly Notices of the Royal Astronomical Society}
\pubyear{2017}
\pagerange{\pageref{firstpage}--\pageref{lastpage}} 

\def\LaTeX{L\kern-.36em\raise.3ex\hbox{a}\kern-.15em
    T\kern-.1667em\lower.7ex\hbox{E}\kern-.125emX}

\begin{document}


\maketitle
\begin{abstract}

The advent of new time domain surveys and the imminent increase in astronomical data expose the shortcomings in traditional time series analysis (such as power spectra analysis) in characterising the abundantly varied, complex and stochastic light curves of Active Galactic Nuclei (AGN). Recent applications of novel methods from non-linear dynamics have shown promise in characterising higher modes of variability and time-scales in AGN. Recurrence analysis in particular can provide complementary information about characteristic time-scales revealed by other methods, as well as probe the nature of the underlying physics in these objects. Recurrence analysis was developed to study the recurrences of dynamical trajectories in phase space, which can be constructed from one-dimensional time series such as light curves. We apply the methods of recurrence analysis to two optical light curves of \textit{Kepler}-monitored AGN. We confirm the detection and period of an optical quasi-periodic oscillation in one AGN, and confirm multiple other time-scales recovered from other methods ranging from 5 days to 60 days in both objects. We detect regions in the light curves that deviate from regularity, provide evidence of determinism and non-linearity in the mechanisms underlying one light curve (KIC 9650712), and determine a linear stochastic process recovers the dominant variability in the other light curve (Zwicky 229--015). We discuss possible underlying processes driving the dynamics of the light curves and their diverse classes of variability. 

\end{abstract}



\section{Introduction} \label{sec:intro}

Most of the extreme radiation output of Active Galactic Nuclei (AGN) originates from the accretion discs surrounding the supermassive black holes at their centers, where gravitational potential energy is converted into heat and viscous dissipation. The disc must transport angular momentum outwards, allowing matter to accrete inwards. 
The radiative flux emitted by the innermost regions of the accretion disc is highly variable across many decades of time, from hours up to months and years (\citealt{Pica1983}, \citealt{Krolik1988}). 
The flux is often assumed to be thermal emission from a geometrically thin disc (\citealt{Pringle1981}, \citealt{Abramowicz2013}), although a variety of possible accretion disc geometries have been proposed (e.g., slim discs, advective-dominated accretion flows, thick dicks; \citealt{Abramowicz2013}). 
The non-periodic and stochastic variability of the radiation emitted from AGN promises to contain a wealth of information regarding the nature of the accretion flow, from the viscous mechanisms generating dissipation \citep{Balbus1991} to the global geometry of the disc \citep{Shakura1973}. 

Changes in the structure of the accretion flow can result in changes in the bulk variability properties that can be observed with photometric monitoring. A typical AGN accretion disc extends to roughly a few tenths of a parsec from the central supermassive black hole \citep{Goodman2003}. Especially at extragalactic distances, direct imaging of such objects is a highly arduous and difficult task \citep{Akiyama2019}, infeasible for the study of temporal changes. With the onset of upcoming sophisticated transient-hunting surveys such as the Large Synoptic Sky Telescope (LSST), the imaging of many thousands of objects observed every night, averaging an estimated 20 TB per 24-hour period\footnote{https://www.lsst.org/scientists/keynumbers}, will surpass a combined decade of imaging data achieved by the Sloan Digital Sky Survey (SDSS)\footnote{The volume of all imaging data collected over a decade by the SDSS-I/II projects published in SDSS DR 7 \citep{Abazajian2009} is approximately 16 TB}\citep{Ivezic2019}. Acquiring corresponding spectroscopic data on the same scale therefore becomes a near impossibility. Astronomers have therefore invested in time series analysis of light curves as the leading probe of dynamical information across the electromagnetic spectrum of accreting, time-varying objects.

There are multiple theoretical processes that have been proposed that potentially explain the rapid and long-term optical variability of AGN. For example, it is theorised that reprocessing of the central X-ray radiation closest to the black hole (\citealt{Krolik1991}, \citealt{Collier1998}, \citealt{Collier2001}), and turbulent or limit cycle thermal processes (\citealt{Kato1998}, \citealt{Shakura1973}) can manifest in oscillations and variability on the short and long term. 

There are several methods to empirically translate the theoretical models of the variability into measurable, time-based quantities. For example, the propagating fluctuations model -- where the fluctuations in local viscosity have been shown to be driven by the magneto-rotational instability \citep{Hogg2016}-- predicts a lognormal distribution of the flux in a light curve and a power spectral density of fluctuations characterised by flicker noise \citep{Lyubarskii1997}. 

There are also statistical explanations for the variability. For example, the observed rms-flux relationship \citep{McHardy2004}, which correlates an increase in luminosity with an increase in variability, led to the relationship between the characteristic time-scale of X-ray variability and black hole mass across many decades of mass \citep{Scaringi2015}. The rms-flux relationship is proposed as a consequence of the multiplicative nature of the propagating fluctuations model \citep{Balbus1998}, however a clear physical interpretation of the rms-flux relationship is still needed. The statistical damped random walk model (DRW, \citealt{Kelly2009}) predicts a power spectral distribution (PSD) with a power-law slope of -2 \citep{MacLeod2010} in which some (unknown) mechanism drives random fluctuations and thus injects stochasticity about the mean flux value into the light curve. The physical source of the random fluctuations and stochasticity is also not well understood.

Lognormal flux distributions, rms-flux relationships, and high-frequency PSD slopes of -2 have been consistently recovered in the X-ray for AGN and X-ray Binaries (XRBs; galactic, stellar-mass black holes) alike (e.g., \citealt{Uttley2001}, \citealt{Edelson2013}). Similarly, in the X-ray bandwidth, a broken power-law model best fits the power spectrum of many XRBs and some AGN, in which the break frequency between the two power-law slopes scales with mass of the object (e.g., \citealt{Uttley2002}, \citealt{Markowitz2003}, \citealt{McHardy2006}).
However, analyses of optical AGN light curves reveal more complex behaviour. Indeed, the rms-flux relationship found in the X-ray and the DRW model does not hold for many of the AGN observed in the optical by \textit{Kepler} (\citealt{Smith2018a}, \citealt{Kasliwal2015a}). Similarly, Moreno et al. (in preparation) finds an array of luminosity and rms-variability relationships for AGN observed by both the Sloan Digital Sky Survey and Catalina Real-Time Transient Survey in addition to a variety of PSD slopes (also uncovered by \citealt{Smith2018a}), which in general are more complex in the optical than in the X-ray. The relationship between the optical and X-ray variability thus remains an open question but could potentially constrain models for the accretion flow.

Commonly used methods for statistical characterisation of light curves, such as the autocorrelation function or PSD, measure only the second-order moments of a distribution. By definition, such methods do not capture the higher-order moments, or traces of non-linearity, non-stationarity and direct probes of the nature of the underlying dynamics (\citealt{Moreno2019}, \citealt{Zbilut2008}). Although, for example, Fourier-based techniques have been the bread and butter of time-domain astronomers and remain some of the most powerful and sophisticated means of characterising dynamical information from light curves, the abundance of discrepancies in empirical PSD-based measures across bandwidths of light (e.g. the rms-flux relationship) and decades of mass (e.g. the presence of a break frequency, and slopes of the PSD) for accreting systems prompts the pursuit of alternative and complementary analyses. Indeed, there has been success in extracting other types of information about the accretion flow by studying the variability of XRBs and AGN using other methods. For example, recurrence analysis (the study of recurrent, non-periodic information) has been used to distinguish between stochastic, periodic and chaotic structures underlying the light curves of six microquasars (XRBs exhibiting some of the properties of a quasar) observed in the X-ray \citep{Sukova2016}. Topological methods derived from group theory \citep{Gilmore1998} related to recurrence analysis were used to positively correlate the non-linear light curve of an X-ray Binary with the chaotic Duffing oscillator \citep{Phillipson2018}. Statistical analyses using CARMA (Continuous-time Auto-Regressive Moving-Average), which critically contain time-ordering information that PSD analyses lack, applied to large AGN surveys have extracted multiple characteristic time-scales in the optical light curves describing the rate at which flux perturbations grow and decay (\citealt{Kasliwal2015b}, \citealt{Kasliwal2017}, \citealt{Moreno2019}). 

The importance of applying alternative methods, which are well established in other non-astrophysical fields such as statistics, economics, or geology, is two-fold. First, we desire a means to more directly probe the source of the time-scales over which various variability properties dominate and identify the mathematical structure of the equations that describe the underlying physics of the variability. For example, we would expect that random flaring in the accretion disc, local fluctuations in the viscosity or accretion rate, or other inherently random-driven processes that do not result in global, coherent structural changes in the accretion disc would lead to a light curve that is well-modelled by a linear, stochastic system. In contrast, the presence of non-linearity in a light curve identified by techniques from non-linear dynamics would provide evidence for a process that is a global instability (e.g. thermal-viscous instabilities or spiral wave modes, e.g. \citealt{Shakura1973}, \citealt{Lightman1974}, \citealt{Wiita1996}), rather than due solely to local fluctuations (e.g. \citealt{Abramowicz1992}, \citealt{McHardy1988}, \citealt{Edelson1999}, \citealt{Poutanen1999}). Although time-scales are important for identifying the possible mechanisms that can exist, correlating specific variability features, such as quasi-periodicity, to a narrower mathematical model (such as non-linearity, or mere determinism), constrains the physical models we construct for accretion disc systems. 

Our secondary motive for employing novel time series analysis techniques is to better prepare for the onset of large datasets from upcoming missions such as LSST. Many efforts are already underway classifying variability features (e.g., based on variability statistics or energy spectra) using automated and fast machine learning methods, such as principle component analysis or self-organising maps (\citealt{Francis1992}, \citealt{Boroson1992}, \citealt{Faisst2019}). We aim to add recurrence properties to the list of variability features that can be classified using machine learning techniques. Additionally, an alternative method to PSD modelling for probing universal variability characteristics will provide complementary constraints on current classification models and enable improved predictions for time sampling and baseline requirements for future surveys of active galaxies and other transients.

In this study we will combine the methods of recurrence analysis and topological non-linear analysis to establish a data-driven approach independent of an assumed model or PSD to extract multiple time-scales of interest and evidence for their underlying mechanisms from the well-sampled, multi-year optical light curves of two canonical AGN monitored by \textit{Kepler}. The recurrence plot is the graphical representation of a 2D matrix which contains information about the recurrences (repetitive but not periodic features) present throughout the light curve, after an embedding into a higher-order space representative of the underlying dynamics. The embedding into a higher-order space, called the phase space, is akin to the transformation performed during singular value decomposition (SVD) or principal component analysis (PCA), where the flux information from the light curve is recast into a mathematically convenient unit-less matrix, while simultaneously maintaining the same topological information that generated the original light curve. The transformation into phase space and the resulting recurrences populate the entries in a matrix, which can be easily plotted into an image called the recurrence plot. The structures in the recurrence plot provide us with topological (dynamical) information about the physical processes that produce the light curve, rather than a merely statistical description.

In Sec.~\ref{sec:sample} we describe the \textit{Kepler} satellite and the resulting data it obtained to construct the two AGN light curves. In Sec.~\ref{sec:results} we use introduce recurrence analysis and use it to identify three characteristic time-scales, all of which were recovered by multiple authors utilising different methods for Zwicky 229--015 (\citealt{Edelson2014}, \citealt{Kasliwal2015a}, \citealt{Kasliwal2015b}, \citealt{Smith2018a}) and one of which was recovered as a low-frequency quasi-periodic oscillation for KIC 9650712 \citep{Smith2018b}. We conclude our results with computing a dynamical invariant, the $K_2$ entropy related to the correlation dimension, as it compares to a series of statistical surrogates using the surrogate data method. The surrogate comparison enables us to possibly distinguish the presence of stochastic and deterministic underlying dynamics in the light curve, which we determine both exist at different horizons and give rise to different time-scales. In Sec.~\ref{sec:conclusions} we explore possible physical mechanisms responsible for the different driving mechanisms that exist in the light curves. We include an appendix with a more detailed overview of recurrence plots and their quantification, collectively called "recurrence analysis," as well as the Surrogate Data method used for establishing significance for our results.

\section{The \textit{Kepler} AGN Sample} \label{sec:sample}

\textit{Kepler} was launched in 2009 and operated for nine years with the scientific objective of exploring the structure and diversity of planetary systems \citep{Borucki2010}. This objective was achieved by searching for repetitive transits in the light curves of extrasolar planetary systems. All stars in the \textit{Kepler} field of view (FOV) were monitored continuously in order to accumulate enough observation time of the transits which only last a fraction of a day. \textit{Kepler} observed $\sim$160,000 exoplanet search target stars with 30 minute sampling for approximately 4 years in the dense target FOV in the region of the sky in the constellations Cygnus and Lyra. Several dozen AGN were discovered within the \textit{Kepler} FOV (e.g., \citealt{Mushotzky2011}, \citealt{Carini2012}). The resulting \textit{Kepler} AGN light curves remain the most well-sampled in the optical bandwidth to date.

\begin{figure*}
\centering
\subfloat{
	\includegraphics[width=0.49\textwidth]{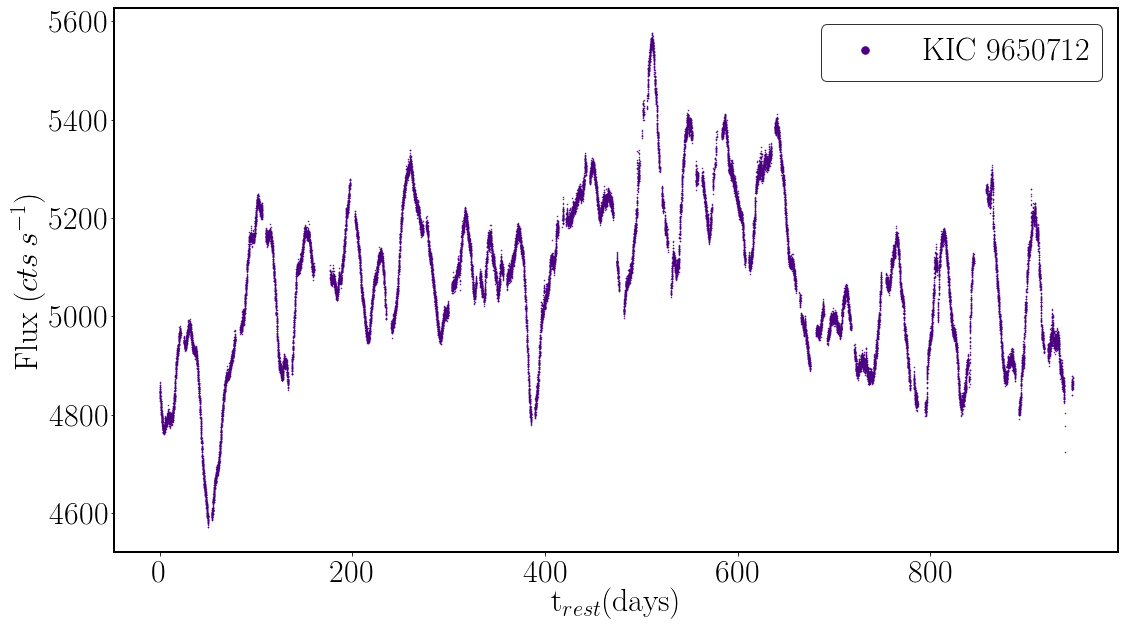}}
\subfloat{
	\includegraphics[width=0.5\textwidth]{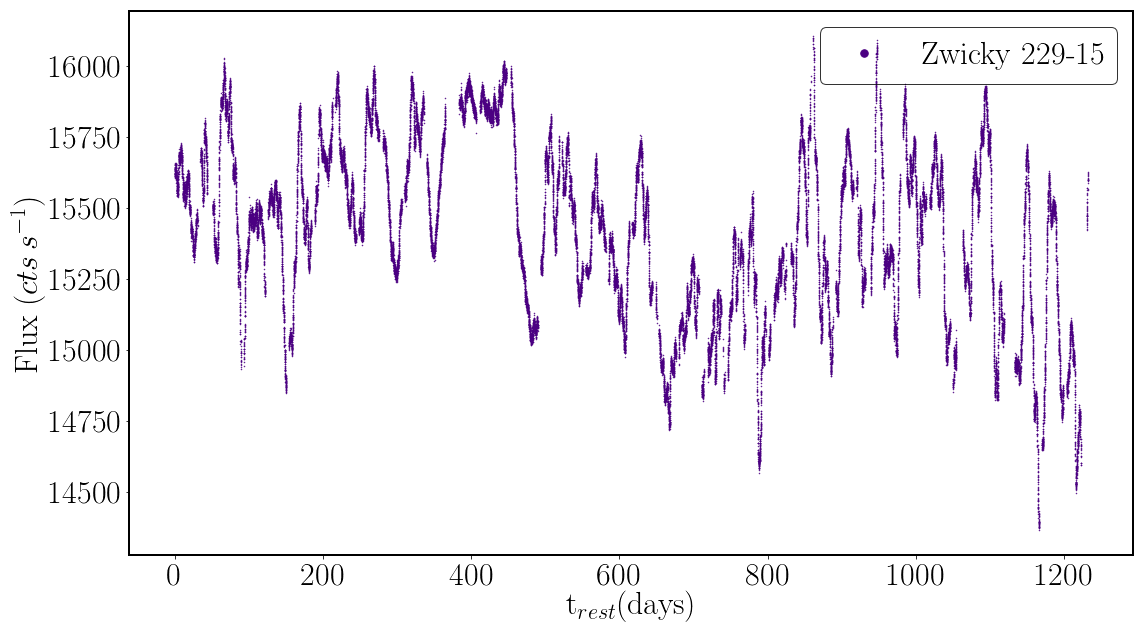}}
\caption{The \textit{Kepler} light curves of (a) KIC 9650712 and (b) Zwicky 229--015, processed by the specialised AGN pipeline by \protect\cite{Smith2018a}. The 30-minute cadence data points are shown in each light curve, with the typical error on each data point approximately 4 flux counts.}
\label{fig:light_curves}
\end{figure*}

A specialised pipeline to construct light curves of a sample of \textit{Kepler}-monitored AGN, selected using infrared photometric selection \citep{Edelson2012}, X-ray selection from KSWAGS (\textit{Kepler}-Swift Active Galaxies and Stars survey; \citealt{Smith2015}), and optical spectroscopy, was developed by \cite{Smith2018a}. The Smith et al. sample contains 21 confirmed AGN light curves with a wide range of accretion rates and black hole masses. We have chosen two long-baseline examples (Fig.~\ref{fig:light_curves}) for which to perform recurrence analysis: the canonical Zwicky 229-015, the longest light curve of an AGN obtained from \textit{Kepler}, and the optical quasi-periodic oscillation candidate, KIC 9650712. A summary of the physical properties of these two AGN, including masses, luminosities, and accretion rates, are listed in Table \ref{tab:objects}. We note that the statistical analyses of \textit{Kepler} AGN light curves may be affected by systematics in the \textit{Kepler} data (see e.g. \citealt{Kasliwal2015b}), which affect the slope of the PSD. Special treatment of the \textit{Kepler} light curves is therefore required (\citealt{Smith2018a} provides an extended discussion on the various systematics and solutions), particularly when considering periodic or quasi-periodic behaviour. When comparing a re-processed light curve of Zw 229--015, calibrated against ground-based observatories to remove systematics introduced particularly at quarterly intervals, to a systematically contaminated light curve, the time-scale associated with a DRW at approximately 26 days remains unchanged \citep{Kasliwal2015b}. This indicates the presence of intrinsic, non-systematic, variability\footnote{The analyses in this paper were performed on both the \cite{Kasliwal2015b} and \cite{Smith2018a} Zw 229--015 \textit{Kepler} light curve and the time-scales recovered using recurrence analysis was the same for both, indicating a non-systematic origin to the variability that is characterised in this study.}.

\begin{table*}
\centering
	\begin{tabular}{ p{1.72cm}p{1.7cm}p{1.72cm}p{0.8cm}p{1.5cm}p{1.25cm}p{2.25cm}p{1.4cm}p{1cm} }
	\hline
		\multicolumn{9}{c}{Physical Properties of the \textit{Kepler} AGN} \\
		\hline
		Object & R.A. & Decl. & z$^a$ &  Kep. Mag.$^b$ & V Mag.$^a$ & log $M_{BH}$ & log $L_{Bol}$$^c$ & $L/L_{Edd}$$^d$ \\
		& & & & & & ($M_\odot$) & ($erg\,\,s^{-1}$)  & \\
		\hline
		KIC 9650712 &  19 29 50.490 & +46 22 23.59  & 0.128 & 16.64 & -21.8 & 8.17$^d$ & 45.62 & 0.226  \\
		KIC 6932990 &  19 05 25.969 & +42 27 40.07 &  0.025  & 11.13 & -19.9 & 6.91$^d$, 7.0$^{+0.19}_{-0.24}$$^e$ & 44.11 & 0.125 \\
		(Zw229--015) &  &  &  &  &  &  &  & \\
	\hline
	\end{tabular}
	\caption[Physical Properties of the Kepler AGN]{\textbf{*Table adapted from \cite{Smith2018a}.}\\
		$^a$Redshift and V-band absolute magnitudes from the VCV Catalog \citep{VCV}\\
		$^b$Generic optical "\textit{Kepler} magnitude" used in the KIC and calculated in \cite{Brown2011}\\
		$^c$Bolometric luminosities calculated by \cite{Runnoe2012}\\
		$^d$Mass and Eddington ratio calculations from \cite{Smith2018a}\\
		$^e$Mass based on H$\beta$ reverberation mapping by \cite{Barth2011}}
	\label{tab:objects}
\end{table*}

\subsection{Zwicky 229--015} \label{subsec:Zw229Data}

There were 7 AGN known to lie in the \textit{Kepler} FOV prior to the satellite's launch in 2009 (\citealt{Mushotzky2011}, \citealt{Barth2011}), one of which has been well-studied: KIC 6932990. Otherwise known as Zw 229--015 (and identified this way throughout the remainder of this paper), this AGN is a radio-quiet Type 1 Seyfert at a redshift of 0.0275 (\citealt{VCV}, or VCV catalog). \cite{Edelson2014} initially found a 5-day characteristic period via power spectrum analysis using the \textit{Kepler} light curve. The \cite{Kasliwal2015a} study recovered a de-correlation time-scale of $\sim$27.5 days extracted from structure function analysis of the \textit{Kepler} light curve and comparison to a DRW model. Using CARMA analysis, the same group \citep{Kasliwal2017} extracted the previously noted time-scale of 5.6 days \citep{Edelson2014} and an additional long-term 67 day time-scale; the CARMA model used was the higher-order damped-harmonic oscillator (DHO) perturbed by a coloured noise process. Finally, \cite{Smith2018a} fit a broken power law to the power spectrum of Zw 229--015 and extracted a characteristic 16.0 day break time-scale which, if we take a note from X-ray studies, should theoretically correlate with the black hole mass. In the same study, the best-fit broken power law extracted a high-frequency PSD slope of -3.4, inconsistent with a DRW model (also found by \citealt{Kasliwal2015a}), affirming the need for higher-order statistical models such as the CARMA DHO model (which can accommodate a variety of fixed or bending PSD slopes; e.g. \citealt{Moreno2019}). 

The multiple studies of Zw 229--015 and resulting characteristic time-scales, each by contrasting methods and differing calibration techniques, have consequently confused any singular explanation for the physical phenomena driving the intrinsic variability in this and, by extrapolation, other similar systems. We therefore seek to apply recurrence analysis to the \textit{Kepler} light curve in order to determine whether we extract the same time-scales as previous studies using different methods, facilitating a more cohesive picture for the accretion process. Furthermore, we seek to develop a method that can add supporting evidence to the relationship between characteristic time-scales and variability features and the intrinsic physical characteristics of the systems (e.g., mass, luminosity, or accretion rate).

\subsection{KIC 9650712} \label{subsec:QPOData}

The other object in the current study, KIC 9650712, was chosen not for its extensive research history but for the recent discovery of an optical quasi-periodic oscillation (QPO) at approximately 44 days \citep{Smith2018b} and its 'interesting' flux distribution, which is neither log-normal nor singularly Gaussian. The origin of QPOs in X-ray binaries remains largely unknown, but it is believed they are associated with the X-rays emitted near the inner edge of the accretion disc. Low-frequency QPOs (on the order of 0.1 to 10 Hz frequencies) have been associated with different spectral states in black hole XRBs \citep{Stiele2013}, an indication that changes in the accretion flow are connected to the manifestation of QPOs. Possible models that could lead to low-frequency QPOs and changing spectral states in XRBs include Lense-Thirring (\citealt{Stella1998}; \citealt{Ingram2010}) or orbital (\citealt{Stella1999}, \citealt{Ingram2012}) precession of the accretion disc, spiral structures in the accretion disc \citep{Tagger1999}, radiation pressure instability \citep{Janiuk2011}, or viscous magneto-acoustic oscillations \citep{Titarchuk2004}. QPOs have primarily been detected in X-rays, with the first X-ray QPO detected in an AGN in 2008 \citep{Gierlinski2008}. KIC 9650712 marks the first detection of a low-frequency QPO in an AGN in the optical bandwidth \citep{Smith2018b}. non-linearity has been confirmed in the X-ray light curves of GRS 1915+105 \citep{Misra2004} and GX 339--4 \citep{Arur2019}, both XRBs, when a QPO was present. We seek to determine whether the same is true of the low-frequency QPO present in the optical light curve of KIC 9650712.

KIC 9650712 is more massive than Zw 229--015 and, notably, intrinsically brighter with a higher Eddington ratio, as detailed in Table ~\ref{tab:objects}. The \cite{Kasliwal2015a} study did perform the same structure function analysis and comparison to a DRW model as with Zw 229--015 and extracted a de-correlation time-scale of $\sim$48 days. Though the \textit{Kepler} light curve used was only 4 quarters of data (compared to the 14 quarters of Zw 229--015 and the 12 used in the current study), \cite{Kasliwal2015a} still noted weak indications of oscillatory behaviour of the shortened light curve. As with Zw 229--015, \cite{Smith2018a} found a broken power-law as the best-fit model for the PSD of KIC 9650712 and extracted a 53 day break time-scale and a high-frequency PSD slope of -2.9. The method of close returns (\citealt{Lathrop1989}, \citealt{Gilmore1993}, \citealt{Mindlin1992}), a subset of recurrence analysis, is capable of extracting unstable periodic orbits in addition to general recurrent behaviour, appropriate for the detection of quasi-periodic signals, which we will utilise to confirm a QPO in KIC 9650712 in Sec.~\ref{subsec: closereturns}.

\section{Results: Characteristic time-scales and Dynamical Invariants} \label{sec:results}

Many traditional time series analysis methods deal with one-dimensional time series and a statistical description of the variability features. In contrast, methods from non-linear dynamics are critically based on the phase space embedding of a time series -- a higher-order space containing the topological information of the time series. The topological structures present in phase space, which most explicitly manifest as recurrences, contain information with direct relationships to the mathematical underpinnings and thus dynamics that generate the time series. 

Recurrences that appear in phase space contain all the information about the dynamics of a system and constitute an alternative, and complete, description of a dynamical system \citep{Robinson2009}. Recurrence Plots (hereafter RPs) were introduced by \cite{Eckmann1987} as a more general means to visualise the recurrences of trajectories embedded in phase space within dynamical systems. RPs are dynamic graphs which provide qualitative and quantitative information about the behaviour of the system of study, particularly indications of stochastic, periodic, or chaotic behaviour. Several of the quantitative measures, particularly those based on diagonal-line features, are mathematically equivalent to a variety of dynamical invariants underlying the observed time series \citep{Webber1994}. For an extensive overview of the history of RPs and their applications, see the seminal review by \cite{Marwan2007}, from which we also draw many of the definitions in this study. Appendix~\ref{sec:recurrence} contains a more detailed discussion of recurrence plots, phase space, and their quantification specific to our analysis.

Following the notation of \cite{Marwan2007}, suppose we have a dynamical system represented by the trajectory ${x_i}$ for $i=1,...,N$ in a $d$-dimensional phase space. Then the recurrence matrix is defined as
\begin{equation}\label{eq:rp_mat1}
\mathbf{R}_{i,j}(\epsilon) = \Theta(\epsilon - ||\vec{x}_i - \vec{x}_j ||) \,\, for \,\, i,j = 1, ... , N,
\end{equation}
where $N$ is the number of time-ordered, measured points ($\vec{x}_i$), $\epsilon$ is a threshold distance, and $\Theta(\cdot)$ is the Heaviside function. For states that persist in an $\epsilon$-neighbourhood, i.e., return to within a threshold distance of a previous state in phase space, the following condition holds:
\begin{equation}\label{eq:condition1}
\vec{x}_i \approx \vec{x}_j \Leftrightarrow \mathbf{R}_{i,j} = 1.
\end{equation}
RPs are thus the graphical representation of the binary recurrence matrix, Eq.~\ref{eq:rp_mat1}, whereby a color represents each entry of the matrix (e.g. a black dot for unity and empty or white for zero). By convention, the axes increase in time rightwards and upwards. The RP is also symmetric about the main diagonal, called the `line of identity' (LOI). 

The distances between two positions in a time series are computed after the time series is embedded in phase space. The phase space is a higher-order space that properly resembles the same topological information as the mechanisms generating the time series (rather than merely the statistical information like the mean or standard deviation of the flux). For theoretical systems in which we know the equations of motion, we can directly embed the time series into a phase space constructed from the derivatives of the system. For experimental data in which we perhaps only have a single observable and no immediate knowledge of the equations of motion, we must construct the phase space. Construction of phase space from experimental data is similar to transforming a dataset via principal component analysis whereby each component of the new vector retains special information about the time series distinct from its scalar form (but does not necessarily retain the units, for example, of the original time series). 

Given that we are dealing with a single observable, the flux, we must construct the multi-dimensional phase space from the one-dimensional light curve. A commonly used method for construction is via the time delay method (\citealt{Whitney1935}, \citealt{Takens1981}), which is an embedding that is a one-to-one mapping to the original attractor that generates the 1D time series without loss of dynamical information \citep{Sauer1991}. Other approaches for constructing the higher-order phase space include independent component analysis \citep{Hyvarinen2001}, singular value decomposition \citep{Broomhead1986}, generalised mutual information \citep{Fraser1989} or numerical derivatives in a differential embedding (if the system is known to be low-dimensional; \citealt{Gilmore1998}). We use the time delay method for phase space reconstruction in our analysis of KIC 9650712 and Zw 229--015, which uses flux values drawn from the light curve separated by the 'time delay' to construct the higher-order vector. For further details of our approach for constructing a phase space embedding based on the time delay method, see Appendix~\ref{subsubsec:PhaseSpace}.

\subsection{The Recurrence Plot: Visualising Structure in the Light Curves}

\begin{figure*}
	\centering
	\includegraphics[width=0.9\textwidth]{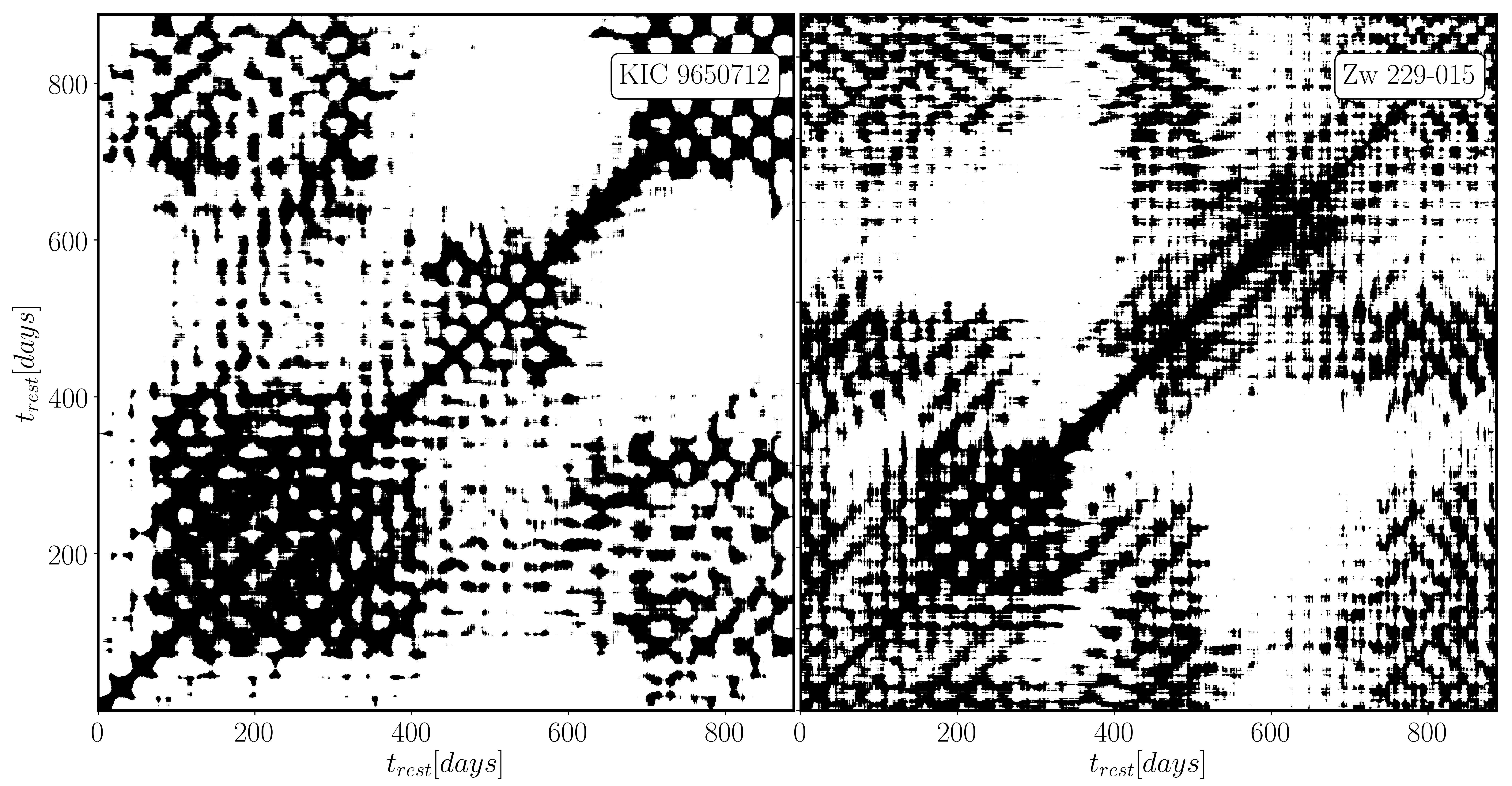}
	\caption{Recurrence Plots of KIC9650712 (left) and Zwicky 229-015 (right) at a recurrence rate (see definition in Sec.~\ref{subsec:RQA}) of 30 per cent; axes are on the same time scale for comparison (though note that the full length of the Zw 229--015 is 200 days longer). A black dot is plotted where the difference in flux at the $x$ and $y$ time positions in the light curve (after embedding in phase space) is less than $\epsilon$.}
	\label{fig:rps_both}
\end{figure*}

\begin{figure*}
\centering
\includegraphics[width=0.95\textwidth]{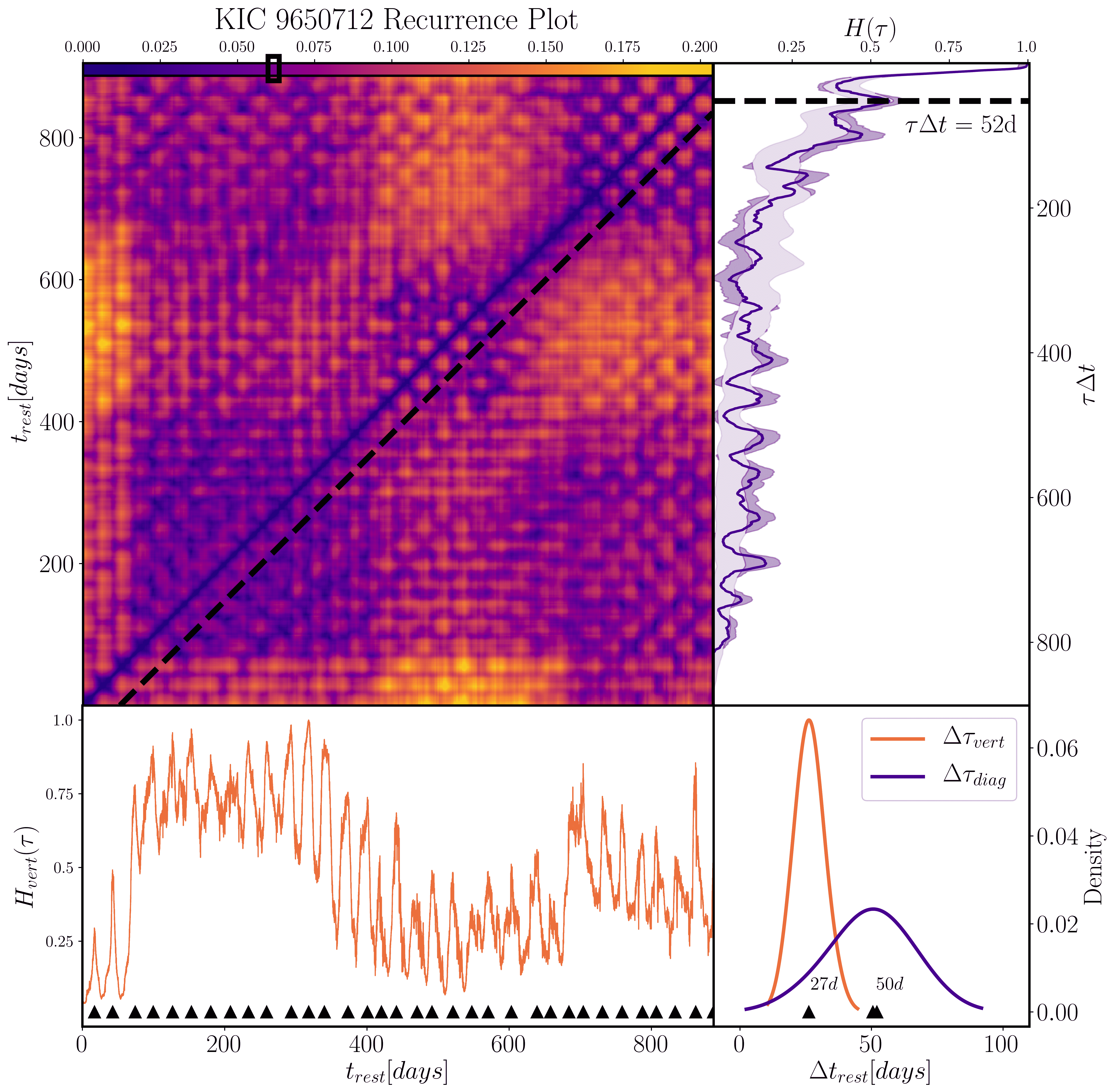}
\caption[RP and Close Returns of KIC 9650712]{Top Left: The un-thresholded Recurrence Plot of KIC 9650712. The colourbar represents the threshold, $\epsilon$, ranging from a corresponding recurrence rate of 1 per cent to 99 per cent, represented in color by purple (dark) to orange (light). \\
Top Right: The close returns histogram, $H(\tau)$ of the RP of KIC 9650712 for a threshold corresponding to RR=30 per cent (marked by the open black rectangle in the colourbar, corresponding to dark purple). The dashed diagonal line from the RP (left) indicates, as an example, the diagonal that was used to compute the first peak in the close returns histogram aligned with the horizontal dashed line (right), at a time delay of $\tau \Delta t = 52\pm2$ days; this period persists with a standard deviation of 9 days for the full histogram. The solid purple (dark) line represents the raw close returns histogram. The regions marked by the purple patches represent the spread in close returns of 100 surrogates with an identical autocorrelation function (ACF) to the data -- the dark (light) purple indicates more (less) than two standard deviations away from the ensemble of ACF surrogates, identifying regions of significance in the data's close returns. \\
Bottom left: The histogram of vertical lines (each column of the RP) at a threshold corresponding to RR=30 per cent. The peaks in the vertical lines histogram are spaced on average by 26 days, with a standard deviation of 4 days, which is approximately the average length of a white patch or line in the standard thresholded RP (represented by the orange patches in the un-thresholded recurrence plot above). \\
Bottom right: The kernel-density estimation (KDE) of the intervals between successive peaks in the vertical lines histogram (orange), with a peak at $26$ days marked by an upward triangle, and close returns histogram (purple), with a peak at $50$ days marked by an upward triangle, calculated with the {\tt pandas} package in Python. The first and most significant time delay of $\tau \Delta t = 52$ days found in the close returns histogram is also marked by an upward triangle, which we note is well aligned with the KDE. }
\label{fig:QPO_rp_closereturns}
\end{figure*}

\begin{figure*}
\centering
\includegraphics[width=0.95\textwidth]{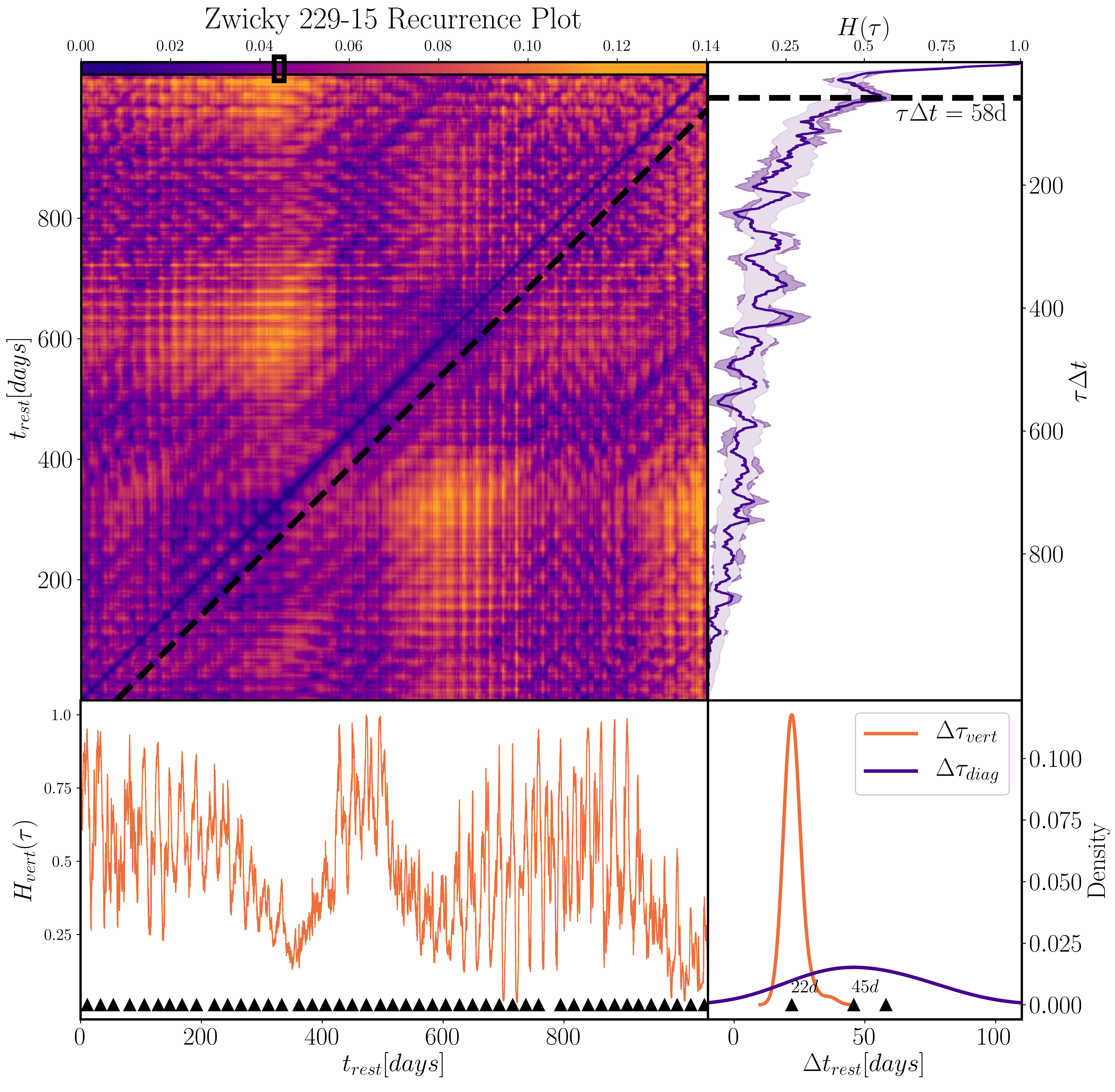}
\caption[RP and Close Returns of Zw 229--015]{The same as Fig.~\ref{fig:QPO_rp_closereturns} but for Zw 229--015. The close returns and vertical line histograms are both computed for a threshold corresponding to 30 per cent, marked by the open black rectangle on the colourbar of the un-thresholded recurrence plot. The average interval between peaks in the vertical lines histogram (bottom left) is $22$ days (standard deviation: 3 days). The KDE of the intervals between successive peaks in the vertical lines histogram (orange, bottom right) also has a peak at $22$ days marked by an upward triangle, and close returns histogram (purple), with a peak at $45$ days marked by an upward triangle, calculated with the {\tt pandas} package in Python. The first and most significant time delay of $\tau \Delta t= 58\pm2$ days found in the close returns histogram (upper right) is also marked by an upward triangle; the peak separations in the close returns histogram produce a standard deviation of 17 days, indicating the instability of the long-term periods, also evident by the deviation in the KDE.}
\label{fig:Zw229_rp_closereturns}
\end{figure*}

An example of the recurrence plots of KIC 9650712 and Zw 229--015 for a threshold corresponding to a recurrence rate of 30 per cent is shown in Fig.~\ref{fig:rps_both}. The light curve of each object is embedded in a higher-order phase space using the time delay method (see Appendix~\ref{subsubsec:PhaseSpace} for details). The distances between every pair of time positions in phase space are computed and if the distance is less than the threshold, then a black dot is plotted (i.e., an entry of one is entered at that matrix position, and zero otherwise). The Euclidean distance metric is used to compute the distance, though any similarity metric could be appropriate. For an in depth discussion of the variety of qualitative features seen for a specific threshold recurrence plot, see Appendix~\ref{subsubsec:PhaseSpace}. For observational data, and for the computation of invariant measures, it is useful to consider the recurrence plot as a function of threshold. We therefore introduce a colourbar representing a range of thresholds for both KIC 950712 and Zw 229--015 displayed in the upper left panel of Fig.~\ref{fig:QPO_rp_closereturns} and Fig.~\ref{fig:Zw229_rp_closereturns}, respectively (\citealt{Zbilut1992}, \citealt{Webber1994}). The colourbar in these figures indicates a range in threshold corresponding to a recurrence point density of $\sim$1 per cent (purple) up to $\sim$99 per cent (orange), which allows for the inspection of the texture of the less recurrent regions of the RP. 

There are multiple features of interest that can provide some insight into the types of behaviour present in the light curves of both objects through a qualitative, visual inspection of the RPs. Both objects have RPs (Fig.~\ref{fig:rps_both}) that display repetitive features vertically (or horizontally, above the LOI) and diagonally (parallel to the LOI) and large white bands and patches (represented by orange in the colourbars of Fig.~\ref{fig:QPO_rp_closereturns} and Fig.~\ref{fig:Zw229_rp_closereturns}). \cite{Marwan2007} notes that periodic and quasi-periodic systems have RPs with diagonally oriented, periodic or quasi-periodic recurrent structures, e.g. diagonal lines and ``checkerboard structures'', the latter of which is most obvious in the KIC 9650712 recurrence plot. In contrast, vertical structures mark time intervals in which the state of the system evolves slowly (or not at all) and is consequently ``trapped'' \citep{Marwan2002}; these features are more obvious in the Zw 229--015 light curve. 

Single, isolated recurrence points reflect both the observational noise and randomness in the light curve while features that fade with increasing distance from the LOI indicate non-stationarity, and large white (orange) bands or patches indicate abrupt changes in the dynamics of the system (e.g., state changes; \citealt{Eckmann1987}). By 'nonstationarity' we mean that the underlying dynamics that produces the light curve are experiencing fluctuations or time-invariance in the state parameters of the equations of motion, detectable over the length of the light curve. In Fig.~\ref{fig:rps_both}, we observe large ``white'' patches (regions that are only considered close for a large threshold; or orange in the colourbars of Fig.~\ref{fig:QPO_rp_closereturns} or  Fig.~\ref{fig:Zw229_rp_closereturns}), indicating possible non-stationarity or dynamical changes in both light curves. We also note, by comparison, the size of recurrent structures (also called ``texture'') in the RPs are different for these two systems, which may indicate stronger higher frequency recurrences or fluctuations in the Zw 229--015 light curve versus KIC 9650712. 

The regions that are globally less recurrent in the un-thresholded RP still contain similar local texture to other regions of the RP (Fig.~\ref{fig:QPO_rp_closereturns} and  Fig.~\ref{fig:Zw229_rp_closereturns}), though less distinct. This may indicate that though the light curve is experiencing a variation in the parameters of the underlying system, the nature of the dynamics driving the variability in the light curve do not cease. We will explore the change in recurrence statistics in the KIC 9650712 recurrence plot in Sec.~\ref{subsubsec: transitions}, where we find the light curve becomes more stochastically driven in the middle of the light curve.

Given the texture of the RPs of both objects, we expect that the light curves contain simultaneous stochastic (or chaotic) and quasi-periodic mechanisms, with the possibility of a dynamics transition particularly notable in the KIC 9650712 RP. The existence of simultaneous periodic and random components has been noted in X-ray binaries (\citealt{Boyd2004}, \citealt{Voges1987}) and the preponderance of one or the other could correlate with intrinsic black hole properties, or specific dominant mechanisms such as a magnetic field (e.g., \citealt{Sukova2016}, \citealt{Ross2017}).

\subsection{Line Features: Quantifying Structure in the Light Curves} \label{subsec: closereturns}

\subsubsection{Diagonal Lines \& Close Returns: Recovering an Optical Quasi-Periodic Oscillation} \label{subsubsec: diagonal}

The structures in the recurrence plot can be quantified, collectively referred to as Recurrence Quantification Analysis, or RQA (see Appendix~\ref{subsec:RQA} for a discussion of the variety of RQA measures in more detail; \citealt{Webber1994}). A variety of RQA measures correlate with specific dynamical invariants, such as Lyapunov exponents (which describe the topological structure of an attractor), dimension, determinism (regions of the time series with high predictability), and laminarity (tendency for regions of a time series to be time-invariant). 

RQA measures can also be computed for each diagonal parallel to the LOI of a RP, thus describing recurrences as a function of time lag in the time series. We define the ``recurrence rate'' of a RP as the percentage of recurrent points with respect to the total size of the recurrence matrix, which is of particular interest when studied as a function of time delay in the time series. That is, for those diagonal lines with distance $\tau$ (number of time-steps or observations) from the LOI, the $\tau$-recurrence rate is defined as
\begin{equation}
RR_{\tau} = \frac{1}{N-\tau} \sum_{i=1}^{N-\tau} \mathbf{R}_{i,i+\tau} = \frac{1}{N-\tau} \sum_{l=1}^{N-\tau} lP_{\tau}(l),
\end{equation}
where $P_{\tau}(l)$ is the number of diagonal lines of length $l$ (in time-steps or observations) on each diagonal parallel to the LOI, offset by a distance $\tau$ (which, when multiplied by the cadence of the light curve, $\Delta t$ can be represented in units of time). The $RR_\tau$ measure can be represented by the so-called ``close returns'' histogram, $H(RR_{\tau})$, which we abbreviate to $H(\tau)$, introduced for quantifying close returns plots (\citealt{Lathrop1989}, \citealt{Gilmore1993}, \citealt{Mindlin1992}). Close returns plots were designed to search for unstable periodic orbits and can be utilised for extracting the periods of strongly recurrent (not necessarily sinusoidal) structure that constitute the `skeleton' of a chaotic attractor (e.g., \citealt{Boyd1994}, \citealt{Phillipson2018}). 

The close returns histogram is conceptually similar to a generalised auto-correlation function (ACF) but describes higher-order correlations between the points of the trajectory in phase space as a function of time delay, $\tau$ \citep{Marwan2007}. A critical difference between $H(\tau)$ and the ACF is the fact that the close returns is drawn from the recurrence plot after embedding in a higher-order space has occurred -- that is, the recurrence peaks in $H(\tau)$ trace the recurrences in the topology of the underlying system, and not merely relationships between delays in the one-dimensional time series. The further advantage of a close returns representation of the data over the (linear) ACF is that it is not an average over an entire sample or single observable but is instead constructed to identify specific, highly correlated segments of data within the time series \citep{Gilmore1993}. This can also be interpreted as the probability that a state recurs to its $\epsilon$-neighbourhood after $\tau$ steps. Indeed, any $\tau$-based RQA measure is capable of finding non-linear similarities in short and non-stationary time series with high noise levels, appropriate for astronomical time series, which surpasses the capabilities of standard ACF techniques \citep{Webber2009}.

We plot the close returns histogram against the un-thresholded RP for KIC 9650712 and Zw 229--015 in Fig.~\ref{fig:QPO_rp_closereturns} and Fig.~\ref{fig:Zw229_rp_closereturns}, respectively. We also construct the close returns of 100 stochastic surrogates (generated as phase-randomised samples from the light curves themselves) that have an identical standard ACF to the original light curve. The stochastic-generated close returns histograms are represented by the spread in light purple. The dark purple patches represent more than two standard deviations away from the ensemble of the ACF surrogates, which we can interpret as regions of significant structure present in the light curves which is not statistically recovered in the ACF.
In other words, the fact there are significant peaks of the close returns histogram for the data with respect to stochastic surrogates which have an identical ACF demonstrates the additional structure that the close returns histogram uncovers versus a standard autocorrelation function.

The first peak in the close returns histogram represents the strongest recurrent period, which we see is also significant, from the diagonal lines. We note that the diagonal lines are responsible for periodic and deterministic structure in the time series. For KIC 9650712 (Fig.~\ref{fig:QPO_rp_closereturns}), the first peak in $H(\tau)$ corresponds to a period of $\tau\Delta t = 52\pm2$ days, apparently consistent with the quasi-periodic oscillation detected by \cite{Smith2018b}. If we then compute the distance between each successive peak in the close returns histogram, we find the 52-day period persists throughout the entirety of the light curve up to many times this fundamental time delay, with a standard deviation of 9 days. We therefore confirm the \cite{Smith2018b} findings that KIC 9650712 contains a quasi-periodic oscillation (QPO), persisting for long-memory times in the light curve.

In contrast, the Zw 229--015 recurrence plot also contains a long-term period of $58\pm2$ days, extracted from the first peak of the close returns histogram (Fig.~\ref{fig:Zw229_rp_closereturns}), but which varies broadly throughout the light curve for large time delays, as can be seen by the flat and wide spread and deviating peak in the kernel density estimation, or KDE, in purple (Fig.~\ref{fig:Zw229_rp_closereturns}, bottom right) of the close returns peak separations. In contrast, the KDE of KIC 9650712 close returns peak separations in Fig.~\ref{fig:QPO_rp_closereturns} is aligned with the QPO detection. We therefore conclude that though a long-term period exists in the Zw 229--015 light curve, its period does not remain stable for multiples of this fundamental period (i.e. for long time delays, the memory in the time series decays rapidly, varying with a standard deviation of 17 days) and the underlying mechanism driving the long-term quasi-periodic fluctuations likely does not dominate the light curve or may not be associated with a deterministic mechanism.

\subsubsection{Vertical Lines \& Recurrence Periods} \label{subsubsec: vertical}

The vertical line structures within the RP result from the intermittent and laminar states of the time series (\citealt{Marwan2002a}; \citealt{Marwan2007}). The average length of a vertical line segment in a RP quantifies the amount of time that the trajectory in a particular state in the underlying system persists, called the ``trapping time,'' $TT$ \citep{Marwan2002}. We can also interpret the trapping time as the length of time that fluctuations in an impulse-response system on average persist. For an accretion disc, these fluctuations originate in the accretion flow on a local scale. Similarly, the time that the trajectory needs to recur to the neighbourhood of a previously visited state, or the time between successive fluctuations, corresponds to a white vertical line in an RP (e.g., the gap between successive states; \citealt{Zou2007}). For example, for periodic motion of period T (perhaps embedded in a noisy signal), we expect a series of uninterrupted diagonal lines separated by a distance T. The vertical distance between successive line segments in the RP, called a ``white'' vertical line, will have a length corresponding to T. The period T is often referred to as the recurrence period, $T_{rec}$ (\citealt{Gao1999}; \citealt{Gao2000}), and is distinct from the dominant phase period, $T_{ph}$, which corresponds to the dominant frequency in the power spectrum of a (possibly noisy, observational) time series (\citealt{Marwan2007}, \citealt{Thiel2003}). 

We estimate $T_{rec}$ through the average white patch length of a RP. The lower left panel in Fig.~\ref{fig:QPO_rp_closereturns} and Fig.~\ref{fig:Zw229_rp_closereturns} is the sum of all vertical line segments in the RP in each column for a specific threshold --- note that this is computed in an identical fashion to the close returns histogram as a sum of the diagonal lines. The peaks in the vertical lines histogram directly pinpoint regions in which we have a high frequency of vertical structure, whereas the distance between successive peaks corresponds to the time delays between the laminar states of the system (the average vertical distance between recurrent patches). The kernel density estimation (KDE) of the peak separations in both histograms is displayed in the bottom right panels of Fig.~\ref{fig:QPO_rp_closereturns} and Fig.~\ref{fig:Zw229_rp_closereturns}, which is much more narrowly isolated for the vertical structure compared to the periods extracted from the close returns. For KIC 9650712 we find the recurrence period to be $26\pm4$ days and for Zw 229--015 we find it to be $22\pm3$ days. We discuss how this time-scale relates to a de-correlation time-scale (e.g., the amount of time for two tangential segments to no longer be correlated), as computed by structure function analysis by \cite{Kasliwal2015a} and extracted from dynamical invariant calculations from the RP (\citealt{Marwan2007}, \citealt{Thiel2003}) in Appendix~\ref{subsubsec: tau_corr}.


\subsection{Distinguishing Deterministic versus Stochastic Mechanisms} \label{sec:K2_RQA}

\subsubsection{K2 Entropy: A Dynamical Invariant Measuring Complexity} \label{subsubsec:entropy}

The most important tracer of regular behaviour, including periodic, quasi-periodic, or deterministic behaviour, results from the existence of long diagonal lines in the RP. The longest diagonal line length in a RP is related to the largest Lyapunov exponent \citep{Eckmann1987}; $L_{max}$ is in fact a good indicator of the presence of determinism \citep{Marwan2007}. However, it is the distribution of diagonal line lengths that is directly related to the correlation entropy (also called the R\'{e}nyi entropy of second order, $K_2$; \citealt{Faure1998}, \citealt{Thiel2003}), which is defined as the lower limit of the sum of the positive Lyapunov exponents \citep{Ruelle1978}. The positive Lyapunov exponents dictate the rate at which trajectories on an attractor diverge for nearby initial conditions, and the negative Lyapunov exponents determine the boundedness of the attractor \citep{Ott}. Chaotic systems contain a positive and finite maximal Lyapunov exponent, $\lambda$, resulting in a $K_2$ entropy that is finite and positive. Perfectly periodic systems have $\lambda=0$ (and thus entropy is also zero), stable fixed points have $\lambda<0$ (and thus entropy is negative or undefined), and noise has $\lambda=\infty$, resulting in an infinite entropy \citep{Kantz}. Thus, the correlation entropy can be used as a discriminating statistic for probing determinism, periodicity, stochasticity, and chaos.

For complex systems with possible quasi-periodic signals, we would expect a small, finite value for the entropy and for deterministic systems, we expect the entropy to be smaller than its dynamics-free surrogates (e.g., statistically generated time series with the same first and second order variability features). The more non-linearity, chaos, or stochasticity present in a system, the larger the value of the entropy. When we compute the correlation entropy of observational data and compare against the entropy calculated from the data's surrogates, we can identify whether dynamical behaviour exists in the data and not in its surrogates and, if so, the nature of the dynamics (e.g., whether it is non-linear or deterministic, or not). In the context of light curves from AGN, the detection of dynamical behaviour that is periodic, deterministic, or non-linear present in the light curve but not in its surrogates, would narrow down the types of mechanisms that generate such behaviour. For example, detection of non-linearity underlying the light curves would rule out models that describe the source of variability as due to, for example, superimposed linear processes (plus uncorrelated noise) of independent active regions in the accretion disc (e.g., \citealt{Terrell1972}; \citealt{Gliozzi2010}).

It has been shown that an estimator for the R\'{e}nyi entropy of the second order, $K_2$, can be obtained directly from an RP \citep{Thiel2004} by
\begin{equation}
K_2(\epsilon, l) = \frac{1}{l\Delta t} \, \log \Big( \frac{1}{N^2} \sum_{t,s=1}^{N} \prod_{m=0}^{l-1} \mathbf{R}_{t+m, s+m} \Big),
\end{equation}
where the quantity within the natural logarithm ($\log$) is the cumulative distribution of diagonal line lengths, $P_{\epsilon}^c(l)$, $l$ is the length of a diagonal line (in number of data points) and $\Delta t$ is the time sampling of the time series. In our case, hourly binning of the light curves was used. When we regard $P_{\epsilon}^c(l)$ as the probability of finding a diagonal line of at least length $l$ in the RP, then the $K_2$ entropy is related by
\begin{equation}\label{eqn:K2}
P_{\epsilon}^c(l) \sim \epsilon^{D_2} \exp^{-K_2(\epsilon) l \Delta t},
\end{equation}
where $D_2$ is the correlation dimension of the system \citep{Thiel2003}. Thus, when we represent $P_{\epsilon}^c(l)$ in a (natural) logarithmic scale versus line length $l$ we obtain a straight line with slope $-K_2(\epsilon)\Delta t$ for large $l$'s, which is an estimate for the correlation entropy. The $K_2$ entropy as a function of thresholds, determined by a $RR$ ranging from 1 per cent to 99 per cent (see e.g. \citealt{Asghari2004}) should be monotonically decreasing and result in a scaling region. The scaling region over a range of thresholds provides a more rigorous estimate of the entropy compared to other methods (e.g., Grassberger-Procaccia method; \citealt{Grassberger1983}) and also accounts for the dynamical and observational noise in the light curves of these physical systems (\citealt{Thiel2002}; \citealt{Thiel2004}). The plateau (scaling region) in the slope of the curves for large $l$ in dependence on $\epsilon$ can be found particularly for chaotic and deterministic systems (\citealt{Marwan2007}, \citealt{Thiel2003}), and is not defined for purely stochastic systems \citep{Thiel2004}. Thus the presence of a scaling region of the entropy with respect to viewing size (threshold) is as important as the value of the entropy for distinguishing between types of dynamical systems (e.g. by the surrogate data method). 

To summarise, the correlation entropy describes the number of possible trajectories that the system can follow within $l$ time steps into the future. That is, the entropy is a proxy for the ``forecasting'' time or horizon of the time series, or how well we can reasonably predict the future for $l\Delta t$ amount of time. From this perspective, for periodic systems, where the largest Lyapunov exponent is zero, the entropy is thus also zero, indicating only one possibility for a future trajectory of the system. For increasing entropy, the possible paths that can be taken into the future increases until, for pure white noise, there are infinite possibilities due to the inherent randomness. 

We note that for well-sampled data, the lines directly above and below the LOI actually represent tangential motion about the LOI rather than distinct orbits. It is thus best practice to exclude this corridor entirely for the determination of dynamical invariants including the entropy \citep{Gao1994}, choosing a width the size of the Theiler window \citep{Theiler1986}, generally comparable to the auto-correlation time. In other words, the entropy will be computed from line lengths that correspond to time-scales longer than approximately 20 days (beyond the de-correlation time-scale, as that found from the vertical lines in Fig.~\ref{fig:QPO_rp_closereturns} and Fig.~\ref{fig:Zw229_rp_closereturns}).

\subsubsection{$K_2$ Entropy: A Comparison to Stochastic Surrogates}

We will use the method of surrogate data (further discussed in Appendix~\ref{subsec:surrogates}; \citealt{Theiler1992}; \citealt{Small2003}), to compare the computation of the $K_2$ entropy for KIC 9650712 and Zw 229--015 against three types of surrogate data sets. Each type of surrogate corresponds to a specific null hypothesis which we compare against using the computation of the entropy. The rejection of a null hypothesis indicates that the light curve is not described by that type of noise process. The three types of surrogates are:
\begin{enumerate}
\item The ``shuffled'' surrogates, which are those that represent temporally independent Gaussian noise (e.g., random drawings from the flux distribution). These surrogates preserve the flux distribution of the original data but destroy the time ordering information \citep{Theiler1992}, and thus represent random observations drawn from the same probability distribution of the data.
\item The ``phase'' surrogates are those that represent linearly correlated Gaussian noise, thereby preserving the autocorrelations (and by extension, the PSD) of the original data, but do not maintain the same flux distribution \citep{Theiler1996}, and thus contain no non-linear determinism. These can be produced by randomising the Fourier phases of the light curve.
\item The final surrogates are generated using the ``IAAFT'' (iterative amplitude adjusted Fourier transform) algorithm, which preserve both the PSD and the flux distribution of the original data \citep{Schreiber1996} and represent static monotonic non-linear transformations of linear noise.
\end{enumerate} 

We will follow the same general approach as introduced by \cite{Small2003}, and brought to astronomical time series by \cite{Sukova2016} and \cite{Asghari2004}, to compute the $K_2$ entropy of the data and their surrogates. Comparison to surrogates that are specifically generated by the light curves themselves means that systematics and noise in the light curves will also, critically, be imposed on to the surrogates. 

\begin{figure*}
\centering
\subfloat{
	\includegraphics[width=0.45\textwidth]{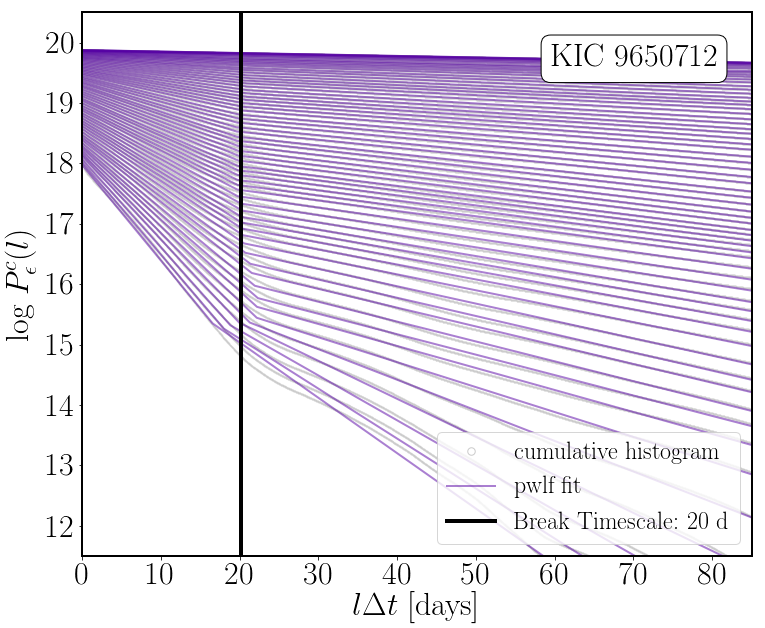}}
\subfloat{
	\includegraphics[width=0.45\textwidth]{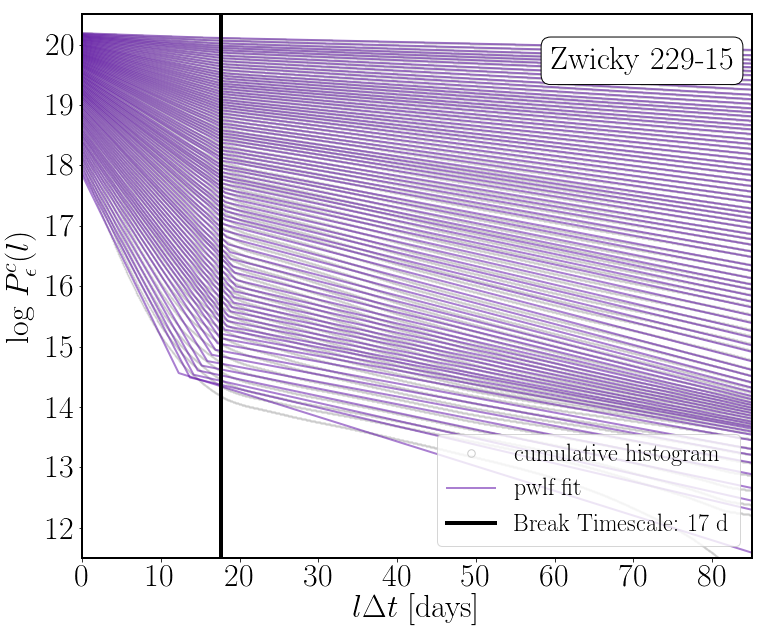}}
\caption[Cumulative Distribution of Diagonal Lines]{The natural logarithm of the cumulative diagonal line lengths histograms as a function of line length for KIC 9650712 (left) and Zw 229--015 (right). A histogram is plotted for each of the 100 thresholds with corresponding recurrence rates ranging between 1 per cent and 99 per cent. The raw cumulative histogram is plotted in light gray data points, and the purple lines are the piece-wise linear fits of the histograms using the Python package, {\tt pwlf}. The solid vertical black line is the average position of the break between the two linear scaling regions, at 20 days for KIC 9650712 and 17 days for Zw 229--015. The slope of each purple line above the break time-scale is used in the computation of the $K_2$ entropy. The histograms are fit out to line lengths corresponding to  approximately 10 per cent the length of the time series \citep{Asghari2004}.}
\label{fig:diagonal_histograms}
\end{figure*}

We utilise three different software packages for a variety of steps in the analysis. These include the publicly available software package {\tt TISEAN}\footnote{https://www.pks.mpg.de/\~{}tisean/} (\citealt{Hegger1999}, \citealt{Schreiber2000}) for the production of surrogates; {\tt PyRQA}\footnote{https://pypi.org/project/PyRQA/} \citep{Rawald2017} for the production of RPs, cumulative diagonal line histograms, and other RQA measures; and finally the Python package {\tt pwlf}\footnote{https://pypi.org/project/pwlf/} (continuous PieceWise Linear Fit) for the linear fitting of the cumulative histograms. In summary, this approach is as follows:

\begin{itemize}
\item We use the procedures {\tt mutual} and {\tt false\_nearest} from {\tt TISEAN} to determine the proper time delay embedding parameters for the construction of the recurrence plots of the light curves (see the Appendix~\ref{subsubsec:PhaseSpace} for a discussion on embedding parameter selection; also used for generating Fig.~\ref{rps_both}-Fig.~\ref{fig:Zw229_rp_closereturns}). We note that though the embedding parameters (time delay and embedding dimension) are required for the production of a RP using {\tt PyRQA}, the results of the computation of the R\'{e}nyi entropy are independent of these parameters \citep{Thiel2004}.
\item Using the {\tt TISEAN} package, we produce 100 surrogates each of the shuffled, phase and IAAFT types. The surrogate generation algorithms are summarised in detail by \cite{Schreiber2000}. Shuffled surrogates are generated through a random shuffling of the original data. The phase surrogates are generated by randomising the Fourier phases of the original data, but maintaining the amplitude of the complex conjugate pairs, and then performing an inverse Fourier transform. The IAAFT surrogates are generated by iteratively filtering towards the correct Fourier amplitudes and rank-ordering to the correct distribution, in an alternating fashion (i.e., an iterated combination of the first two algorithms).
\item For each of the original data of KIC 9650712 and Zw 229--015 and all of their surrogates we produce a recurrence plot for 100 thresholds ranging from $\epsilon_{min}$ to $\epsilon_{max}$, corresponding to RR=1 per cent to 99 per cent, using the {\tt PyRQA} package. The colourbars in Fig.~\ref{fig:QPO_rp_closereturns} and Fig.~\ref{fig:Zw229_rp_closereturns} cover the range of all thresholds used.
\item For each of the 3 types of surrogates and the original light curve of each object, we produce a cumulative distribution of diagonal line lengths, $P_{\epsilon}^c(l)$, for every threshold and use the {\tt pwlf} package to fit the linear regions in the $\log \, P_{\epsilon}^c(l)$ versus diagonal line length $l$ plot. Fig.~\ref{fig:diagonal_histograms} shows the logarithmic plot of $P_{\epsilon}^c(l)$ and associated line fits for both objects, KIC 9650712 and Zw 229--015, for all thresholds.
\item With the line fits and resulting slopes of the cumulative histograms in hand, we compute the $K_2$ entropy as a function of threshold, $\epsilon$, for all time series. \cite{Asghari2004} determined that the $K_2$ entropy should be fit by 3 ``clusters'', where the region with the flattest slope represents the optimal estimate for the entropy (recall, for deterministic and non-linear systems, there is a plateau in the entropy for a range of neighbourhood sizes, but a plateau may not exist if the system is linear stochastic, for example). We use the {\tt pwlf} package again to fit 3 regions of the $K_2$ entropy and choose the smallest slope region as our best estimate for the entropy for every object and each of its surrogates. That is, we use the same threshold range to compute $K_2$ for all surrogates as we do for the original data, for consistency.
\item We compute the significance of the $K_2$ entropy against each of the surrogate types as a function of $\epsilon$ in two ways. First, we use the standard rank-order test used for most statistical tests in the surrogate data method (see Appendix~\ref{subsec:surrogates}; \citealt{Theiler1992}) to compute the significance of the deviation of the entropy of the data from each of its surrogates as a function of threshold. We select a residual probability $\alpha$ of a false rejection, corresponding to a significance $(1-\alpha) \times 100\%$ for a generated $M=K/\alpha -1$ surrogates. The probability that the data by coincidence has one of the smallest values is exactly $\alpha$. For our given 100 surrogates, a 95 per cent confidence level that the null hypothesis is rejected would correspond to our data representing one of the 5 smallest values of the entropy for a given threshold, as we expect purely stochastic systems to have higher entropy. Secondly, given the distribution of the $K_2$ entropy as a function of threshold is important for distinguishing non-linearity or determinism (i.e. the existence of a plateau), we use the 2-sample Kolmogorov-Smirnov (KS) test \citep{Smirnov1939} using {\tt SciPy} to compare the empirical distributions of the entropy calculations for all thresholds of the data versus each of its surrogates, where we would expect our data to have a significantly different distribution from all of its surrogates if it contains determinism or non-linearity.
\end{itemize}

\begin{figure*}
\centering
\includegraphics[width=0.95\textwidth]{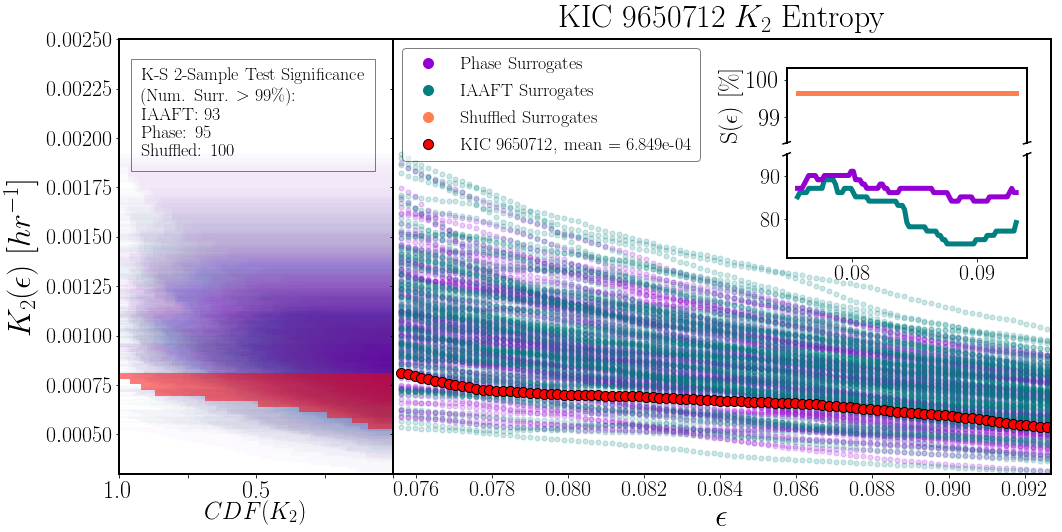}
\caption[Entropy of KIC 9650712]{The $K_2$ entropy of KIC 9650712 as a function of threshold, $\epsilon$, for the original data (larger red-filled circles) and the three surrogate types: shuffled in orange, which is so different from the data and other surrogates that is is out of the field of view of these plots; phase surrogates are in purple; and IAAFT surrogates are in teal. The mean entropy for KIC 9650712 is $6.849 \times 10^{-4}$ \textrm{hr}$^{-1}$. The inset in the plot shows the significance of the entropy calculation for the data as a function of threshold for each of the three surrogate types: for each column (threshold), a rank-order test is computed for the data against the surrogates, e.g. if the data (red) is one of the five smallest values including surrogates, it is significant against the surrogates at a 95 per cent confidence level. The shuffled surrogates are highly significant, the phase surrogates peak above a 90 per cent significance, and the IAAFT surrogates peak at 90 per cent significance. We note the plateau in the $K_2$ entropy for KIC 9650712. The left panel is the cumulative histogram of the entropy for all thresholds for the data (red) and each of the surrogates (orange, purple, and teal as in the right panel). For each surrogate, a two-sample KS test is performed against the data cumulative histogram: 12 surrogates total are indistinguishable from the data by the KS test at a 99.9 per cent confidence level -- 7 from IAAFT surrogates, 5 from phase surrogates, and none from the shuffled surrogates.}
\label{fig:K2_QPO}
\end{figure*}

\begin{figure*}
\centering
\includegraphics[width=0.95\textwidth]{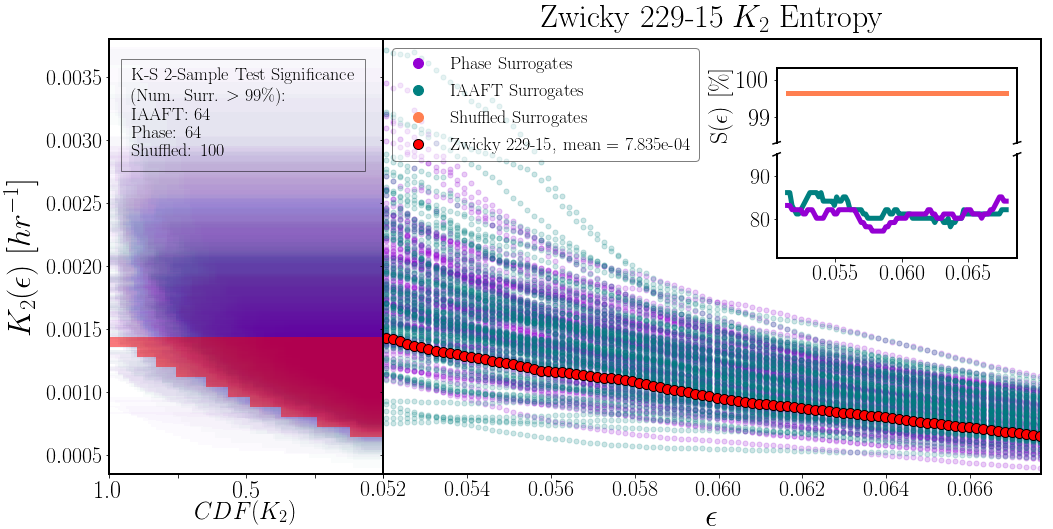}
\caption[Entropy of Zw 229--015]{The same as Fig.~\ref{fig:K2_QPO}. The $K_2$ entropy of Zw 229--015 as a function of threshold, $\epsilon$, for the original data (larger red-filled circles) and the three surrogate types: shuffled in orange, which is so different from the data and other surrogates that is is out of the field of view of these plots; phase surrogates in purple; and IAAFT surrogates in teal. The mean entropy for Zw 229--015 is $1.035 \times 10^{-3}$ \textrm{hr}$^{-1}$, systematically higher than KIC 9650712. The significance of the entropy calculation for the data as a function of threshold for each of the three surrogate types is highly significant for the shuffled surrogates, but well below 90 per cent for both the phase and IAAFT surrogate types. There is no evidence for a plateau in the $K_2$ entropy. When computing the two-sample KS test of the cumulative histogram of the entropy for the data against each of the surrogates, many are indistinguishable from the data at the 99.9 per cent confidence level.}
\label{fig:K2_Zwicky}
\end{figure*}

The mean $K_2$ entropy in the full threshold range is $6.849 \times 10^{-4}$ $\textrm{hr}^{-1}$ for KIC 9650712 and $1.035 \times10^{-3}$ $\textrm{hr}^{-1}$ for Zw 229--015. We reiterate that the absolute value of the entropy for each object in and of itself has little meaning without comparison to surrogates, since we are dealing with observational systems with inherent noise and systematics (versus theoretical dynamical systems with well known dimension). The significance against the three types of surrogates is higher for KIC 9650712 than is for Zw 229--015. For KIC 9650712, we see the rank-order test of the entropy as a function of threshold reveals above 90 per cent confidence level of significance against the phase and IAAFT surrogate types for small thresholds and above a 99 per cent confidence level against the shuffled surrogates. When performing the 2-sample KS test of the distribution of entropy against all of the surrogate types, 12 total surrogates (none from the shuffled surrogates) were coincidentally similar to the data out of all 300 surrogates --- i.e., the difference in distributions constituted a less than 95 per cent level of significance that the null hypothesis is false for only 12 surrogates. We conclude that the entropy for KIC 9650712 is modestly systematically lower than the surrogates, but strongly indicates the presence of determinism. For Zw 229--015, only the shuffled surrogates have a higher than 95 per cent confidence level of significance for both the rank-order test and the 2-sample KS test; the phase and IAAFT surrogates never reach a 90 per cent confidence level for low thresholds in the rank-order test, and their distributions in entropy are not significantly different from the data when compared via the 2-sample KS test. 

The results of the surrogate data analysis include:
\begin{enumerate}
\item KIC 9650712 as compared to Zw 229--015 contains more regular (or deterministic) behaviour. This is evident in the fact that there appears to be a plateau in the $K_2(\epsilon)$ plot of KIC 9650712 (Fig.~\ref{fig:K2_QPO}) but not in that for Zw 229--015 (Fig.~\ref{fig:K2_Zwicky}), which is what we would expect from deterministic or chaotic systems, but not of linear stochastic ones. 
\item The rejection of the null hypothesis from the shuffled surrogates for both objects is highly significant. This means we can, unsurprisingly, rule out a temporally independent Gaussian process as a major contributor to the observed variability in both systems. At a minimum, the variability has significant correlations.
\item The null hypothesis of a linear correlated stochastic process (from the phase surrogates) can be likely ruled out for KIC 9650712 -- a Gaussian process does not give rise to the variability -- but is not significant enough for Zw 229--015. The same is true for a possibly non-linearly rescaled linear stochastic process (from the IAAFT surrogates, where the flux distribution is preserved in addition to the PSD) --- non-linearity in the noise response of the KIC 9650712 light curve also does not dominantly contribute to the variability. This means KIC 9650712 either contains non-linear structure in one of the underlying physical mechanisms, or there is significant variation of the state parameters of the underlying system over the length of the light curve (e.g., a possible dynamical state change). Our analysis does not distinguish between non-linearity and non-stationarity of this kind.
\end{enumerate}

The main results of this analysis are that there appears to be an underlying deterministic mechanism in the KIC 9650712 light curve driving variability on time-scales beyond 20 days (based on the comparison of the entropy to surrogate data; Fig.~\ref{fig:K2_QPO}), including the quasi-periodic oscillation (Fig.~\ref{fig:QPO_rp_closereturns}), with the presence of possible non-stationarity or non-linearity in the underlying mechanism. Meanwhile, the variability of Zw 229--015 is indistinguishable from a linear stochastic process, when the entropy is compared to surrogate data (Fig.~\ref{fig:K2_Zwicky}). In the case of Zw 229--015, this means that the light curve can be well modelled by a typical stochastic process, such as the ARMA or CARMA(2,1) damped harmonic oscillator \citep{Moreno2019} or similar, in which the linear autocorrelations recover a majority of the observed variability. In contrast, the KIC 9650712 light curve contains simultaneous stochastic (e.g., on time-scales less than the de-correlation time) and deterministic periodic components (e.g., possibly associated with the QPO) and thus we should look towards alternative models, such as non-linear oscillators, to characterise its variability.

It must be pointed out that we chose three specific null hypotheses for which to test the light curves of KIC 9650712 and Zw 229--015. When we reject a null hypothesis, as we did for all surrogate types for KIC 9650712 (though, modestly) and for just the shuffled surrogates of Zw 229--015, it is important we distinguish that these specific hypotheses do not represent the data for the specific discriminating statistic that we used (the correlation entropy). Similarly, for Zw 229--015, failure to reject a null hypothesis (e.g., the phase and IAAFT surrogates were not rejected) does not necessarily mean that the null hypothesis is true, only that the the correlation entropy failed to be a probe of the differences between the null hypotheses and the data. 

\subsubsection{RQA: Determinism, time-scales, and transitions} \label{subsubsec: transitions}

\begin{figure*}
\centering
\subfloat{
	\includegraphics[width=0.45\textwidth]{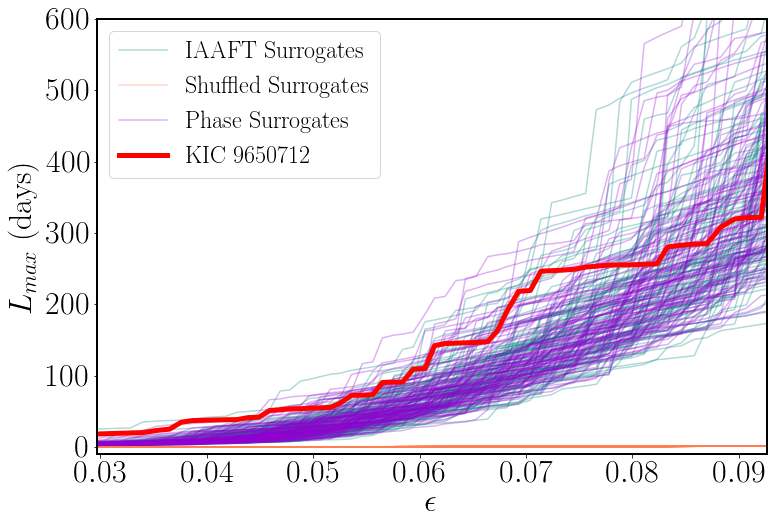}}
\subfloat{
	\includegraphics[width=0.45\textwidth]{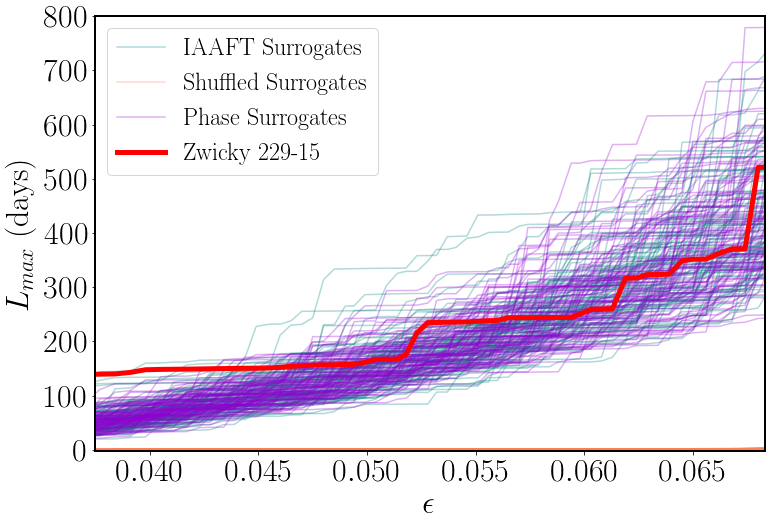}}
\caption{The length of the longest diagonal line found in the RP of KIC 9650712 (left) and Zw 229--015 (right) as a function of threshold, in a solid red line, against the three surrogate types (shuffled surrogates in orange, phase surrogates in purple, and IAAFT surrogates in teal). The longest diagonal line remains significant against all surrogates for a wide range of thresholds for KIC 9650712, and for only the smallest of thresholds for Zw 229--015. If the longest diagonal line length is significant against various surrogate types, then there is a strong likelihood that the underlying dynamics are deterministic. An identical calculation was performed for multiple variability states of six XRBs by \protect\cite{Sukova2016}, in which $L_{max}$ also distinguished deterministic structure in certain variability states, and did not for stochastic variability states.}
\label{fig:Lmax} 
\end{figure*}

We can confirm the presence of determinism by studying the maximum length of a diagonal line in the RP of each object with respect to each of its surrogates as a function of threshold. As referenced in Sec.~\ref{subsubsec:entropy}, the longest diagonal line that is present in an RP is an indicator for deterministic structure, a calculation from the RP that is computationally much faster than the $K_2$ entropy and thus can be used for larger samples of objects. The entropy calculation is the most rigorous comparison to surrogates, but does require well-sampled data (on the order of 10,000 to 30,000 data points, such as from \textit{Kepler}) in order to recover a scaling region as a function of threshold \citep{Eckmann1992}. In contrast, the longest diagonal line, or any other recurrence statistic, can be computed with fewer data points, on the order of $\sim 1,000$ or less \citep{Marwan2007}. In Fig.~\ref{fig:Lmax} we see that $L_{max}$ is significant against all surrogates for a wide range of thresholds for KIC 9650712 ($L_{max}$ is systematically longer), but not for Zw 229--015. We therefore confirm the presence of determinism in the KIC 9650712 light curve using a different discriminating statistic from the entropy, and do not conclude determinism is evidenced in the Zw 229--015 light curve. It is also important to point out that the \cite{Sukova2016} study found that systems with QPOs (e.g., XRBs in the $\rho$, or ``heartbeat,'' state) were similarly significant against all surrogate types and deduced to contain non-linear dynamics from their light curves. 

We can explore other recurrence statistics from the RPs of KIC 9650712 and Zw 229--015 related to characteristic time-scales and indications of transitions in the dynamics. We have recovered two characteristic time-scales from the recurrence plots of KIC 9650712 and Zw 229--015: a quasi-periodic long-term time-scale on the order of 50 days or more from the close returns (Fig.~\ref{fig:QPO_rp_closereturns} and Fig.~\ref{fig:Zw229_rp_closereturns}), and a de-correlation time-scale from the frequency of vertical lines (also derived from the cumulative distribution of diagonal lines, Fig.~\ref{fig:diagonal_histograms}). Both these time-scales are related to how often a state will recur and the probability of the occurrence of a particular state as a function of time lag. The third, and shortest, time-scale that can be recovered from recurrence analysis relates to how long a state will persist, which can be estimated by the average length of a diagonal line, $L_{avg}$, and the average length of a vertical line, $TT$, called the trapping time.  KIC 9650712 has a $TT = 4.9$ days and an average diagonal line length of $L_{avg} = 2.98$ days. Zw 229--015 has a $TT= 4.8$ days and an average diagonal line length of $L_{avg} = 2.7$ days. We note that a $\sim 5$ day characteristic time-scale was recovered from the Zw 229--015 \textit{Kepler} light curve via power spectrum analysis \citep{Edelson2014} and from structure function analysis \citep{Kasliwal2015a}, the latter of which indicates that the time-scales at this length may be related to the average persistence time of an impulse fluctuation in an $AR$-type process. Indeed, if we consider how the average line lengths evolve over time, where we can compute $L_{avg}$ and $TT$ in a sliding window across the entire light curve, both $L_{avg}$ and $TT$ vary between 2 days and 20 days, with shorter lines in the middle of the light curve. Similarly, if we divide the KIC 9650712 into three segments, compute the recurrence plot and from it the $K_2$ entropy for each segment separately, we find the entropy is highest (e.g. more noise-like) in the middle of the light curve ($K_2 = -2.78 \times 10^{-3}$ \textrm{hr}$^{-1}$) compared to the beginning and end of the light curve ($K_2 = -1.56 \times 10^{-3}$ \textrm{hr}$^{-1}$ and $K_2 = -1.72 \times 10^{-3}$ \textrm{hr}$^{-1}$, respectively). The change in length of the average line may therefore be a quantitative measure for the change in texture in the recurrence plot and thus an analog for the more computationally intensive entropy.  An investigation into windowed recurrence analysis of a set of known state-transitioning X-ray binaries with a comparison to spectra is the subject of a subsequent paper.

\section{Conclusions} \label{sec:conclusions}

The qualitative information that a recurrence plot can provide is in itself useful for distinguishing a variety of time series. However, the structure of recurrences, when quantified, not only indicates the underlying dynamical system but it has been shown that recurrences also contain all the information about the dynamics of a system and constitute an alternative, and complete, description of a dynamical system \citep{Robinson2009}. We determine the structure of recurrences using the Recurrence Plot for two Active Galactic Nuclei monitored in the optical by the \textit{Kepler} satellite: we first confirm characteristic time-scales of interest identified by other methods, which verifies the validity of using RPs for AGN analysis; and we secondly find evidence for low-dimensional determinism in one object (KIC 9650712), and primarily stochastic realisations of underlying processes in the other (Zw 229--015).

In summary, we find three characteristic time-scales derived from the recurrence plot, which we correlate to three different processes:
\begin{itemize}
\item Both objects contain a long-term time-scale of $52\pm2$ days for KIC 9650712 and $58\pm2$ days for Zwicky 229--015. In the KIC 9650712 light curve, this period persists for many multiples of the fundamental period and is consistent with the previously detected optical low-frequency quasi-periodic oscillation \citep{Smith2018b}. 
Furthermore, the organisation of the diagonal lines in the recurrence plot, which give rise to the $\sim 50$ day time-scale, can be quantified by the correlation entropy. 
When the entropy is compared to a series of surrogate data, we see evidence that the long-term behaviour in the KIC 9650712 light curve is likely driven by a low-dimensional deterministic, possibly non-linear and/or non-stationary, process. In contrast, in the Zw 229--015 light curve, the period does not persist, but instead decays rapidly with time, and the mechanism determining the long-term variability is likely indistinguishable from a stochastic process.
\item A de-correlation time-scale of $26\pm4$ days for KIC 9650712 and $22\pm3$ days for Zw 229--015 was detected from the frequency of vertical line structures in the RP, which corresponds to the average amount of time between successive variability states in the light curve and indicates the amount of time that must pass before two points in the light curve are no longer correlated \citep{Kasliwal2015a}.
\item We determine the average length of a diagonal or vertical line in the recurrence plot is $\sim$2-5 days for both KIC 9650712 and Zw 229--015 and corresponds to the average length of time that a specific, localised variability state will persist. This can be interpreted as the average amount of time that a localised fluctuation persists in a statistical impulse-response model, as described by autoregressive processes (e.g., \citealt{Moreno2019}). Furthermore, the lengths of the average diagonal and vertical lines in the RP as a function of time are probes of periodic-chaos/chaos-periodic transitions, chaos-chaos transitions, and changes in laminar states \citep{Marwan2007}.
\end{itemize}

We conclude that recurrence analysis is capable of recovering time-scales probed by other methods, such as from the power spectrum, autocorrelation function, structure function, or stochastic modelling (\citealt{Edelson2014}, \citealt{Kasliwal2015a}, \citealt{Smith2018a}, \citealt{Smith2018b}, \citealt{Moreno2019}). Furthermore, recurrence analysis is capable of providing evidence for the nature of the underlying processes that produce the light curve related to these characteristic time-scales. We compute an estimate for the dynamical invariant of the R\'{e}nyi entropy of second order, $K_2$ (also called the correlation entropy), directly from the recurrence plot of both KIC 9650712 and Zw 229--015 and compare the results to three types of surrogate data, each representing a different stochastic null hypothesis, using the surrogate data method of \cite{Theiler1992}. We determine that the KIC 9650712 light curve is likely driven by a deterministic process, with possible non-linearity or non-stationarity, on the order of many tens of days, while the Zw 229--015 light curve may be well-modelled by a linear, stochastic process in which the linear autocorrelations recover the majority of the observed variability. Though this is a case study of only two objects, we hypothesise that the determinism in the KIC 9650712 light curve is related to the presence of the quasi-periodic oscillation (QPO), previously detected by \cite{Smith2018b}. 

Since the development of the surrogate data method, there have been advancements in more rigorous and sophisticated null hypotheses and testing procedures (e.g., as summarised by \citealt{Lancaster2018}), which may be more suitable for analysing the full 21 \textit{Kepler} AGN sample from \cite{Smith2018a}. For example, Moreno \textrm{et al.} (paper forthcoming) finds that AGN observed by SDSS and the CRTS could be well-modelled by two classes of CARMA processes: one is the damped random walk, and the other is a stochastic damped harmonic oscillator. Rejection (or failure to reject) of a null hypothesis based on autoregressive moving average processes would corroborate the results from Moreno et al. and possibly further suggest two classes of the underlying physical process of the light curve. We also point out here that when we say `non-stationarity', we refer to the time variance of the parameters of the underlying system or of dynamical transitions over the course of the light curve, which can be significant due to size effects. The methods we have utilised in this paper provide evidence for non-linearity, but we have not determined whether the source of the non-linearity is due to non-stationarity of this kind and thus it is possible that a state transition in the AGN light curve was captured over this time period. A windowed recurrence plot analysis would help illuminate whether state transitions have occurred (e.g., as discussed in Sec.~\ref{subsubsec: transitions}).

QPOs have been uncovered in the X-ray light curves of both X-ray Binaries (XRBs) and AGN; the QPO signal in KIC 9650712 represents the first optical detection in an AGN, and its connection to X-ray variability remains unclear. The lack of a confirmation of the rms-flux relationship in the \textit{Kepler} AGN light curves \citep{Smith2018a}, an empirical phenomenon previously detected in the X-ray of AGN, suggests that the propagating fluctuations model for the accretion disc may not be a consistent model for these observations in the optical and similarly the optical variability may not solely be due to reprocessing of the X-ray light from the innermost accretion disc or hot corona (e.g. there may be instabilities arising directly in the optical regions of the accretion disc). Given the deterministic nature of KIC 9650712 and the presence of a QPO, random flaring in the accretion disc or localised fluctuations in the accretion rate are unlikely to be the dominant source of the variability on the order of many days in the KIC 9650712 light curve. Instead, mechanisms capable of producing limit-cycle behaviour and entering a non-linear regime on a global scale must be the primary source of variability at these time-scales for KIC 9650712. Furthermore, that the correlation entropy with respect to stochastic surrogates is more significant for a QPO source is consistent with the results found for six microquasars using the same method \citep{Sukova2016}, which contained stronger evidence for non-linearity particularly in the strongly QPO-like variability states of the XRBs. This suggests that there may be a common accretion mechanism in both XRBs and AGN that leads to QPO behaviour.

The detection of non-linearity alongside QPO signals has occurred in XRBs and microquasars at both short-term time-scales (e.g., seconds and sub-seconds, \citealt{Sukova2016}) and long-term time-scales (e.g., many days, \citealt{Phillipson2018}). The apparent self-similarity across many decades of time, a hallmark of non-linear and chaotic systems, adds support to the prospect of a non-linear physical mechanism driving variability associated with quasi-periodic behaviour. However, some of the processes proposed for QPOs in XRBs would likely not be detected in AGN on the order of many days as in this study. For example, the radiation pressure instability would occur on the order of thousands of years or more \citep{Janiuk2011}, as would precession of the accretion disc connected to jet precession \citep{Lu1990} or the radiation-driven instability (\citealt{Petterson1977}; \citealt{Pringle1996}; \citealt{Armitage1997}). However, the disc precession models are typically based on the assumption that information is transported through diffusion \citep{Pringle1997}; if instead the propagation of a warp in the accretion disc was transported via wave-like processes, then the speed at which they propagate would be closer to the sound speed, corresponding to variability times much shorter than the viscous diffusion time, a feasible option for $\sim$50 d time-scales. 

Another possible disc precession model is the magnetically-driven instability (\citealt{Aly1980}, \citealt{Lai2003}), originating from the Bardeen-Petterson effect due to frame-dragging at the innermost edge of the accretion disc linking the spin of the central black hole to the magnetic field of the accretion disc (i.e., Lense-Thirring precession, \citealt{Bardeen1975}). In this case, the optical variability and QPOs would be intrinsically tied to, and possibly phase-shifted from, the X-ray QPOs \citep{Veledina2013}. A multi-wavelength study of AGN containing QPOs would be required to confirm this scenario, especially given the unclear manifestations of the rms-flux relationship in the optical.  

We reiterate that the two-object sample in this study is clearly not sufficient to make any claims about which of these processes gives rise to the QPO signal in KIC 9650712 and further study of a large sample of AGN light curves, ideally multi-wavelength, is required. We merely hypothesise that the time-scale of the QPO and its moderate significance as a deterministic process with possibly non-linear origin indicates that the mechanisms producing the optical quasi-periodicity may either be due to an inner accretion disc process that propagates outwards (possibly distinct from the rms-flux relationship and propagating fluctuations model), or one that originates in the optical region of the accretion disc and is transported through wave-like processes (in order to occur on days-months periods). In either case the driving mechanism must be capable of operating in a non-linear regime. A sampling of the parameter space of various recurrence quantities with respect to physical characteristics of an ensemble of AGN and XRBs may help illuminate dependencies on physical characteristics of the systems, such as accretion rate and luminosity, and is the subject of a subsequent paper.

\section*{Acknowledgements}
This study is based on work fully supported by the National Aeronautics and Space Administration (NASA) under Grant Numbers NNX16AT15H and 80NSSC19K1291 issued through the NASA Education Minority University Research Education Project (MUREP) through the NASA Harriett G. Jenkins Graduate Fellowship activity. This research made use of data reduced and provided by Krista Lynne Smith while at the University of Maryland, originating from the \textit{Kepler} satellite. R.A.N. also acknowledges Krista Lynne Smith for helpful discussions about the \textit{Kepler} AGN and their properties, Jackeline Moreno for sharing expertise on time series analysis and CARMA modelling, and Stephen McMillan for general feedback on the mathematical underpinnings of recurrence analysis and the surrogate data method. M.S.W. acknowledges support from the Ambrose Mondell Foundation during sabbatical leave at the Institute for Advanced Study. 

\bibliography{References}

\begin{thebibliography}{}
\makeatletter
\relax
\def\mn@urlcharsother{\let\do\@makeother \do\$\do\&\do\#\do\^\do\_\do\%\do\~}
\def\mn@doi{\begingroup\mn@urlcharsother \@ifnextchar [ {\mn@doi@}
  {\mn@doi@[]}}
\def\mn@doi@[#1]#2{\def\@tempa{#1}\ifx\@tempa\@empty \href
  {http://dx.doi.org/#2} {doi:#2}\else \href {http://dx.doi.org/#2} {#1}\fi
  \endgroup}
\def\mn@eprint#1#2{\mn@eprint@#1:#2::\@nil}
\def\mn@eprint@arXiv#1{\href {http://arxiv.org/abs/#1} {{\tt arXiv:#1}}}
\def\mn@eprint@dblp#1{\href {http://dblp.uni-trier.de/rec/bibtex/#1.xml}
  {dblp:#1}}
\def\mn@eprint@#1:#2:#3:#4\@nil{\def\@tempa {#1}\def\@tempb {#2}\def\@tempc
  {#3}\ifx \@tempc \@empty \let \@tempc \@tempb \let \@tempb \@tempa \fi \ifx
  \@tempb \@empty \def\@tempb {arXiv}\fi \@ifundefined
  {mn@eprint@\@tempb}{\@tempb:\@tempc}{\expandafter \expandafter \csname
  mn@eprint@\@tempb\endcsname \expandafter{\@tempc}}}

\bibitem[\protect\citeauthoryear{Abazajian et~al.,}{Abazajian
  et~al.}{2009}]{Abazajian2009}
Abazajian K.~N.,  et~al., 2009, \mn@doi [ApJSS] {10.1088/0067-0049/182/2/543},
  182

\bibitem[\protect\citeauthoryear{Abramowicz \& Fragile}{Abramowicz \&
  Fragile}{2013}]{Abramowicz2013}
Abramowicz M.~A.,  Fragile P.~C.,  2013, {Foundations of black hole accretion
  disk theory}, \mn@doi{10.12942/lrr-2013-1}

\bibitem[\protect\citeauthoryear{Abramowicz, Lanza, Spiegel  \&
  Szuszkiewicz}{Abramowicz et~al.}{1992}]{Abramowicz1992}
Abramowicz M.~A.,  Lanza A.,  Spiegel E.~A.,   Szuszkiewicz E.,  1992, Letters
  to Nature, 356, 41

\bibitem[\protect\citeauthoryear{Akiyama et~al.,}{Akiyama
  et~al.}{2019}]{Akiyama2019}
Akiyama K.,  et~al., 2019, \mn@doi [ApJ] {10.3847/2041-8213/ab0ec7}, 875, L1

\bibitem[\protect\citeauthoryear{Aly}{Aly}{1980}]{Aly1980}
Aly J.,  1980, A{\&}A, 86, 192

\bibitem[\protect\citeauthoryear{Anishchenko, Vadivasova, Okrokvertskhov  \&
  Strelkova}{Anishchenko et~al.}{2003}]{Anishchenko2003}
Anishchenko V.~S.,  Vadivasova T.~E.,  Okrokvertskhov G.~A.,   Strelkova G.~I.,
   2003, in Physica A. pp 199--212, \mn@doi{10.1016/S0378-4371(03)00199-7}

\bibitem[\protect\citeauthoryear{Armitage \& Pringle}{Armitage \&
  Pringle}{1997}]{Armitage1997}
Armitage P.~J.,  Pringle J.~E.,  1997, \mn@doi [ApJ] {10.1086/310907}

\bibitem[\protect\citeauthoryear{Arur \& Maccarone}{Arur \&
  Maccarone}{2019}]{Arur2019}
Arur K.,  Maccarone T.~J.,  2019, \mn@doi [MNRAS] {10.1093/mnras/stz1052}, 486,
  3451

\bibitem[\protect\citeauthoryear{Asghari et~al.,}{Asghari
  et~al.}{2004}]{Asghari2004}
Asghari N.,  et~al., 2004, \mn@doi [A{\&}A] {10.1051/0004-6361:20040390}, 426,
  353

\bibitem[\protect\citeauthoryear{Babaei, Zarghami, Sedighikamal,
  Sotudeh-Gharebagh  \& Mostoufi}{Babaei et~al.}{2014}]{Babaei2014}
Babaei B.,  Zarghami R.,  Sedighikamal H.,  Sotudeh-Gharebagh R.,   Mostoufi
  N.,  2014, \mn@doi [Physica A] {10.1016/j.physa.2013.10.016}, 395, 112

\bibitem[\protect\citeauthoryear{Balbus \& Hawley}{Balbus \&
  Hawley}{1991}]{Balbus1991}
Balbus S.~A.,  Hawley J.~F.,  1991, ApJ, 376, 214

\bibitem[\protect\citeauthoryear{Balbus \& Hawley}{Balbus \&
  Hawley}{1998}]{Balbus1998}
Balbus S.~A.,  Hawley J.~F.,  1998, \mn@doi [Rev. of Mod. Phys.]
  {10.1103/revmodphys.70.1}, 70, 1

\bibitem[\protect\citeauthoryear{Bardeen \& Petterson}{Bardeen \&
  Petterson}{1975}]{Bardeen1975}
Bardeen J.~M.,  Petterson J.~A.,  1975, \mn@doi [ApJL] {10.1086/181711}, 195,
  L65

\bibitem[\protect\citeauthoryear{Barth et~al.,}{Barth et~al.}{2011}]{Barth2011}
Barth A.~J.,  et~al., 2011, \mn@doi [ApJ] {10.1088/0004-637X/732/2/121}, 732

\bibitem[\protect\citeauthoryear{Boroson \& Green}{Boroson \&
  Green}{1992}]{Boroson1992}
Boroson T.~A.,  Green R.~F.,  1992, \mn@doi [ApJSS] {10.1086/191661}, 80, 109

\bibitem[\protect\citeauthoryear{Borucki et~al.,}{Borucki
  et~al.}{2010}]{Borucki2010}
Borucki W.~J.,  et~al., 2010, \mn@doi [Science] {10.1126/science.1185402}, 327,
  977

\bibitem[\protect\citeauthoryear{Boyd \& Smale}{Boyd \& Smale}{2004}]{Boyd2004}
Boyd P.~T.,  Smale A.~P.,  2004, ApJ, 612, 1006

\bibitem[\protect\citeauthoryear{Boyd, Mindlin, Gilmore  \& Solari}{Boyd
  et~al.}{1994}]{Boyd1994}
Boyd P.~T.,  Mindlin G.~B.,  Gilmore R.,   Solari H.~G.,  1994, ApJ, 431, 425

\bibitem[\protect\citeauthoryear{Broomhead \& King}{Broomhead \&
  King}{1986}]{Broomhead1986}
Broomhead D.~S.,  King G.~P.,  1986, \mn@doi [Physica D]
  {10.1016/0167-2789(86)90031-X}, 20, 217

\bibitem[\protect\citeauthoryear{Brown, Latham, Everett  \& Esquerdo}{Brown
  et~al.}{2011}]{Brown2011}
Brown T.~M.,  Latham D.~W.,  Everett M.~E.,   Esquerdo G.~A.,  2011, \mn@doi
  [AJ] {10.1088/0004-6256/142/4/112}, 142

\bibitem[\protect\citeauthoryear{Carini \& Ryle}{Carini \&
  Ryle}{2012}]{Carini2012}
Carini M.~T.,  Ryle W.~T.,  2012, \mn@doi [ApJ] {10.1088/0004-637X/749/1/70},
  749

\bibitem[\protect\citeauthoryear{Collier \& Peterson}{Collier \&
  Peterson}{2001}]{Collier2001}
Collier S.,  Peterson B.~M.,  2001, ApJ, 555, 775

\bibitem[\protect\citeauthoryear{Collier et~al.,}{Collier
  et~al.}{1998}]{Collier1998}
Collier S.~J.,  et~al., 1998, AJ, 500, 162

\bibitem[\protect\citeauthoryear{Eckmann \& Ruelle}{Eckmann \&
  Ruelle}{1992}]{Eckmann1992}
Eckmann J.~P.,  Ruelle D.,  1992, \mn@doi [Physica D]
  {10.1016/0167-2789(92)90023-G}

\bibitem[\protect\citeauthoryear{Eckmann, {Oliffson Kamphorst}  \&
  Ruelle}{Eckmann et~al.}{1987}]{Eckmann1987}
Eckmann J.-P.,  {Oliffson Kamphorst} S.,   Ruelle D.,  1987, Europhysics
  Letters, 4, 973

\bibitem[\protect\citeauthoryear{Edelson \& Goddard}{Edelson \&
  Goddard}{1999}]{Edelson1999}
Edelson R.,  Goddard N.,  1999, ApJ, 514, 682

\bibitem[\protect\citeauthoryear{Edelson \& Malkan}{Edelson \&
  Malkan}{2012}]{Edelson2012}
Edelson R.,  Malkan M.,  2012, \mn@doi [ApJ] {10.1088/0004-637X/751/1/52}, 751

\bibitem[\protect\citeauthoryear{Edelson, Mushotzky, Vaughan, Scargle, Gandhi,
  Malkan  \& Baumgartner}{Edelson et~al.}{2013}]{Edelson2013}
Edelson R.,  Mushotzky R.,  Vaughan S.,  Scargle J.,  Gandhi P.,  Malkan M.,
  Baumgartner W.,  2013, \mn@doi [ApJ] {10.1088/0004-637X/766/1/16}, 766, 16

\bibitem[\protect\citeauthoryear{Edelson, Vaughan, Malkan, Kelly, Smith, Boyd
  \& Mushotzky}{Edelson et~al.}{2014}]{Edelson2014}
Edelson R.,  Vaughan S.,  Malkan M.,  Kelly B.,  Smith K.,  Boyd P.,
  Mushotzky R.,  2014, \mn@doi [ApJ] {10.1088/0004-637X/795/1/2}, 795, 2

\bibitem[\protect\citeauthoryear{Faisst, Prakash, Capak  \& Lee}{Faisst
  et~al.}{2019}]{Faisst2019}
Faisst A.~L.,  Prakash A.,  Capak P.~L.,   Lee B.,  2019, \mn@doi [ApJ]
  {10.3847/2041-8213/ab3581}, 881, L9

\bibitem[\protect\citeauthoryear{Faure \& Korn}{Faure \&
  Korn}{1998}]{Faure1998}
Faure P.,  Korn H.,  1998, Physica D, 122, 265

\bibitem[\protect\citeauthoryear{Francis, Hewett, Foltz  \& Chaffee}{Francis
  et~al.}{1992}]{Francis1992}
Francis P.~J.,  Hewett P.~C.,  Foltz C.~B.,   Chaffee F.~H.,  1992, ApJ, 398,
  476

\bibitem[\protect\citeauthoryear{Fraser}{Fraser}{1989}]{Fraser1989}
Fraser A.~M.,  1989, in Abraham N.,  ed., , Measures of Complexity and Chaos.
Plenum Press, New York, pp 117--119, \mn@doi{10.1007/978-1-4757-0623-9_11}

\bibitem[\protect\citeauthoryear{Fraser \& Swinney}{Fraser \&
  Swinney}{1986}]{Fraser1986}
Fraser A.~M.,  Swinney H.~L.,  1986, Phys. Rev. A, 33

\bibitem[\protect\citeauthoryear{Gao}{Gao}{1999}]{Gao1999}
Gao J.~B.,  1999, PRL, 83

\bibitem[\protect\citeauthoryear{Gao \& Cai}{Gao \& Cai}{2000}]{Gao2000}
Gao J.,  Cai H.,  2000, \mn@doi [Phys. Lett. A]
  {10.1016/S0375-9601(00)00304-2}, 270, 75

\bibitem[\protect\citeauthoryear{Gao \& Zheng}{Gao \& Zheng}{1994}]{Gao1994}
Gao J.,  Zheng Z.,  1994, \mn@doi [Phys. Rev. E] {10.1103/PhysRevE.49.3807},
  49, 3807

\bibitem[\protect\citeauthoryear{Gierli{\'{n}}ski, Middleton, Ward  \&
  Done}{Gierli{\'{n}}ski et~al.}{2008}]{Gierlinski2008}
Gierli{\'{n}}ski M.,  Middleton M.,  Ward M.,   Done C.,  2008, \mn@doi
  [Nature] {10.1038/nature07277}, 455, 369

\bibitem[\protect\citeauthoryear{Gilmore}{Gilmore}{1993}]{Gilmore1993}
Gilmore C.~G.,  1993, J. of Econ. Behavior and Org., 22, 209

\bibitem[\protect\citeauthoryear{Gilmore}{Gilmore}{1998}]{Gilmore1998}
Gilmore R.,  1998, Rev. of Mod. Phys., 70, 1455

\bibitem[\protect\citeauthoryear{Gliozzi, R{\"{a}}th, Papadakis  \&
  Reig}{Gliozzi et~al.}{2010}]{Gliozzi2010}
Gliozzi M.,  R{\"{a}}th C.,  Papadakis I.~E.,   Reig P.,  2010, \mn@doi
  [A{\&}A] {10.1051/0004-6361/200912948}, 512, A21

\bibitem[\protect\citeauthoryear{Goodman}{Goodman}{2003}]{Goodman2003}
Goodman J.,  2003, MNRAS, 339, 937

\bibitem[\protect\citeauthoryear{Grassberger}{Grassberger}{1983}]{Grassberger1983}
Grassberger P.,  1983, \mn@doi [Phys. Lett. A] {10.1016/0375-9601(83)90753-3},
  97, 227

\bibitem[\protect\citeauthoryear{Hegger, Kantz  \& Schreiber}{Hegger
  et~al.}{1999}]{Hegger1999}
Hegger R.,  Kantz H.,   Schreiber T.,  1999, \mn@doi [Chaos]
  {10.1063/1.166424}, 9

\bibitem[\protect\citeauthoryear{Hilborn}{Hilborn}{2001}]{Hilborn}
Hilborn R.,  2001, {Chaos and Nonlinear Dynamics: An Introduction for
  Scientists and Engineers}.
Oxford University Press

\bibitem[\protect\citeauthoryear{Hogg \& Reynolds}{Hogg \&
  Reynolds}{2016}]{Hogg2016}
Hogg J.~D.,  Reynolds C.~S.,  2016, \mn@doi [ApJ] {10.3847/0004-637x/826/1/40},
  826, 40

\bibitem[\protect\citeauthoryear{Hyv{\"{a}}rinen, Hoyer  \&
  Inki}{Hyv{\"{a}}rinen et~al.}{2001}]{Hyvarinen2001}
Hyv{\"{a}}rinen A.,  Hoyer P.~O.,   Inki M.,  2001, \mn@doi [Neural
  Computation] {10.1162/089976601750264992}, 13, 1527

\bibitem[\protect\citeauthoryear{Ingram \& Done}{Ingram \&
  Done}{2010}]{Ingram2010}
Ingram A.,  Done C.,  2010, \mn@doi [MNRAS] {10.1111/j.1365-2966.2010.16614.x},
  405, 2447

\bibitem[\protect\citeauthoryear{Ingram \& Done}{Ingram \&
  Done}{2012}]{Ingram2012}
Ingram A.,  Done C.,  2012, \mn@doi [MNRAS] {10.1111/j.1365-2966.2012.21907.x},
  427, 934

\bibitem[\protect\citeauthoryear{Ivezi{\'{c}} et~al.,}{Ivezi{\'{c}}
  et~al.}{2019}]{Ivezic2019}
Ivezi{\'{c}} {\v{Z}}.,  et~al., 2019, \mn@doi [ApJ] {10.3847/1538-4357/ab042c},
  873, 111

\bibitem[\protect\citeauthoryear{Janiuk \& Czerny}{Janiuk \&
  Czerny}{2011}]{Janiuk2011}
Janiuk A.,  Czerny B.,  2011, \mn@doi [MNRAS]
  {10.1111/j.1365-2966.2011.18544.x}, 414, 2186

\bibitem[\protect\citeauthoryear{Kantz \& Schreiber}{Kantz \&
  Schreiber}{2004}]{Kantz}
Kantz H.,  Schreiber T.,  2004, {Nonlinear Time Series Analysis}.
Cambridge University Press

\bibitem[\protect\citeauthoryear{Kasliwal, Vogeley  \& Richards}{Kasliwal
  et~al.}{2015a}]{Kasliwal2015a}
Kasliwal V.~P.,  Vogeley M.~S.,   Richards G.~T.,  2015a, \mn@doi [MNRAS]
  {10.1093/mnras/stv1230}, 451, 4328

\bibitem[\protect\citeauthoryear{Kasliwal, Vogeley, Richards, Williams  \&
  Carini}{Kasliwal et~al.}{2015b}]{Kasliwal2015b}
Kasliwal V.~P.,  Vogeley M.~S.,  Richards G.~T.,  Williams J.,   Carini M.~T.,
  2015b, \mn@doi [MNRAS] {10.1093/mnras/stv1797}, 453, 2075

\bibitem[\protect\citeauthoryear{Kasliwal, Vogeley  \& Richards}{Kasliwal
  et~al.}{2017}]{Kasliwal2017}
Kasliwal V.~P.,  Vogeley M.~S.,   Richards G.~T.,  2017, \mn@doi [MNRAS]
  {10.1093/mnras/stx1420}, 470, 3027

\bibitem[\protect\citeauthoryear{Kato}{Kato}{1998}]{Kato1998}
Kato S.,  ed. 1998, {Black-hole accretion disks}

\bibitem[\protect\citeauthoryear{Kelly, Bechtold  \& Siemiginowska}{Kelly
  et~al.}{2009}]{Kelly2009}
Kelly B.~C.,  Bechtold J.,   Siemiginowska A.,  2009, \mn@doi [ApJ]
  {10.1088/0004-637X/698/1/895}, 698, 895

\bibitem[\protect\citeauthoryear{Kennel, Brown  \& Abarbanel}{Kennel
  et~al.}{1992}]{Kennel1992}
Kennel M.~B.,  Brown R.,   Abarbanel H. D.~I.,  1992, Phys. Rev. A, 45

\bibitem[\protect\citeauthoryear{Krolik \& Begelman}{Krolik \&
  Begelman}{1988}]{Krolik1988}
Krolik J.~H.,  Begelman M.~C.,  1988, \mn@doi [ApJ] {10.1086/166414}, 329, 702

\bibitem[\protect\citeauthoryear{Krolik, Horne, Kallman, Malkan, Edelson  \&
  Kriss}{Krolik et~al.}{1991}]{Krolik1991}
Krolik J.~H.,  Horne K.,  Kallman T.~R.,  Malkan M.~A.,  Edelson R.~A.,   Kriss
  G.~A.,  1991, ApJ, 371, 541

\bibitem[\protect\citeauthoryear{Lai}{Lai}{2003}]{Lai2003}
Lai D.,  2003, \mn@doi [ApJL] {10.1086/377163}, 591, L119

\bibitem[\protect\citeauthoryear{Lancaster, Iatsenko, Pidde, Ticcinelli  \&
  Stefanovska}{Lancaster et~al.}{2018}]{Lancaster2018}
Lancaster G.,  Iatsenko D.,  Pidde A.,  Ticcinelli V.,   Stefanovska A.,  2018,
  {Surrogate data for hypothesis testing of physical systems},
  \mn@doi{10.1016/j.physrep.2018.06.001}

\bibitem[\protect\citeauthoryear{Lathrop \& Kostelich}{Lathrop \&
  Kostelich}{1989}]{Lathrop1989}
Lathrop D.~P.,  Kostelich E.~J.,  1989, Phys. Rev. A, 40

\bibitem[\protect\citeauthoryear{Lightman \& Eardley}{Lightman \&
  Eardley}{1974}]{Lightman1974}
Lightman A.~P.,  Eardley D.~M.,  1974, ApJ, 187, 1

\bibitem[\protect\citeauthoryear{Lu}{Lu}{1990}]{Lu1990}
Lu J.,  1990, A{\&}A, 229, 424

\bibitem[\protect\citeauthoryear{Lyubarskii}{Lyubarskii}{1997}]{Lyubarskii1997}
Lyubarskii Y.~E.,  1997, \mn@doi [MNRAS] {10.1093/mnras/292.3.679}, 292, 679

\bibitem[\protect\citeauthoryear{MacLeod et~al.,}{MacLeod
  et~al.}{2010}]{MacLeod2010}
MacLeod C.~L.,  et~al., 2010, \mn@doi [ApJ] {10.1088/0004-637X/721/2/1014},
  721, 1014

\bibitem[\protect\citeauthoryear{Markowitz et~al.,}{Markowitz
  et~al.}{2003}]{Markowitz2003}
Markowitz A.,  et~al., 2003, ApJ, 593, 96

\bibitem[\protect\citeauthoryear{Marwan, Thiel  \& Nowaczyk}{Marwan
  et~al.}{2002a}]{Marwan2002}
Marwan N.,  Thiel M.,   Nowaczyk N.~R.,  2002a, \mn@doi [Nonlinear Processes in
  Geophysics] {10.5194/npg-9-325-2002}, 9, 325

\bibitem[\protect\citeauthoryear{Marwan, Wessel, Meyerfeldt, Schirdewan  \&
  Kurths}{Marwan et~al.}{2002b}]{Marwan2002a}
Marwan N.,  Wessel N.,  Meyerfeldt U.,  Schirdewan A.,   Kurths J.,  2002b,
  \mn@doi [Phys. Rev. E] {10.1103/PhysRevE.66.026702}, 66

\bibitem[\protect\citeauthoryear{Marwan, {Carmen Romano}, Thiel  \&
  Kurths}{Marwan et~al.}{2007}]{Marwan2007}
Marwan N.,  {Carmen Romano} M.,  Thiel M.,   Kurths J.,  2007, {Recurrence
  plots for the analysis of complex systems},
  \mn@doi{10.1016/j.physrep.2006.11.001}

\bibitem[\protect\citeauthoryear{McHardy, Papadakis, Uttley, Page  \&
  Mason}{McHardy et~al.}{2004}]{McHardy2004}
McHardy I.~M.,  Papadakis I.~E.,  Uttley P.,  Page M.~J.,   Mason K.~O.,  2004,
  \mn@doi [MNRAS] {10.1111/j.1365-2966.2004.07376.x}, 348, 783

\bibitem[\protect\citeauthoryear{McHardy, Koerding, Knigge, Uttley  \&
  Fender}{McHardy et~al.}{2006}]{McHardy2006}
McHardy I.~M.,  Koerding E.,  Knigge C.,  Uttley P.,   Fender R.~P.,  2006,
  \mn@doi [Nature] {10.1038/nature05389}, 444, 730

\bibitem[\protect\citeauthoryear{Mchardy}{Mchardy}{1988}]{McHardy1988}
Mchardy I.,  1988, Memorie della Societa Astronomica Italiana, 59, 239

\bibitem[\protect\citeauthoryear{Mindlin \& Gilmore}{Mindlin \&
  Gilmore}{1992}]{Mindlin1992}
Mindlin G.~B.,  Gilmore R.,  1992, Physica D, 58, 229

\bibitem[\protect\citeauthoryear{Misra, Harikrishnan, Mukhopadhyay, Ambika  \&
  Kembhavi}{Misra et~al.}{2004}]{Misra2004}
Misra R.,  Harikrishnan K.~P.,  Mukhopadhyay B.,  Ambika G.,   Kembhavi A.~K.,
  2004, ApJ, 609, 313

\bibitem[\protect\citeauthoryear{Moreno, Vogeley, Richards  \& Yu}{Moreno
  et~al.}{2019}]{Moreno2019}
Moreno J.,  Vogeley M.~S.,  Richards G.~T.,   Yu W.,  2019, \mn@doi [PASP]
  {10.1088/1538-3873/ab1597}, 131, 63001

\bibitem[\protect\citeauthoryear{Mushotzky, Edelson, Baumgartner  \&
  Gandhi}{Mushotzky et~al.}{2011}]{Mushotzky2011}
Mushotzky R.~F.,  Edelson R.,  Baumgartner W.,   Gandhi P.,  2011, \mn@doi
  [ApJL] {10.1088/2041-8205/743/1/L12}, 743, 6

\bibitem[\protect\citeauthoryear{Ott}{Ott}{2002}]{Ott}
Ott E.,  2002, {Chaos in Dynamical Systems}, 2 edn.
Cambridge University Press, \mn@doi{10.1017/CBO9780511803260}

\bibitem[\protect\citeauthoryear{Petterson}{Petterson}{1977}]{Petterson1977}
Petterson J.~A.,  1977, \mn@doi [ApJ] {10.1086/155280}

\bibitem[\protect\citeauthoryear{Phillipson, Boyd  \& Smale}{Phillipson
  et~al.}{2018}]{Phillipson2018}
Phillipson R.~A.,  Boyd P.~T.,   Smale A.~P.,  2018, \mn@doi [MNRAS]
  {10.1093/MNRAS/STY970}, 477, 5220

\bibitem[\protect\citeauthoryear{Pica \& Smith}{Pica \& Smith}{1983}]{Pica1983}
Pica A.~J.,  Smith A.~G.,  1983, ApJ, 272, 11

\bibitem[\protect\citeauthoryear{Poincar{\'{e}}}{Poincar{\'{e}}}{1890}]{Poincare}
Poincar{\'{e}} H.,  1890, Acta mathematica, 13, 1

\bibitem[\protect\citeauthoryear{Pompe}{Pompe}{1993}]{Pompe1993}
Pompe B.,  1993, J. of Stat. Phys., 73

\bibitem[\protect\citeauthoryear{Poutanen \& Fabian}{Poutanen \&
  Fabian}{1999}]{Poutanen1999}
Poutanen J.,  Fabian A.~C.,  1999, MNRAS, 306, 31

\bibitem[\protect\citeauthoryear{Pringle}{Pringle}{1981}]{Pringle1981}
Pringle J.~E.,  1981, Ann. Rev. of Astronomy and Astrophysics, 19, 137

\bibitem[\protect\citeauthoryear{Pringle}{Pringle}{1996}]{Pringle1996}
Pringle J.~E.,  1996, MNRAS, 281, 357

\bibitem[\protect\citeauthoryear{Pringle}{Pringle}{1997}]{Pringle1997}
Pringle J.~E.,  1997, MNRAS, 292, 136

\bibitem[\protect\citeauthoryear{Rawald, Sips  \& Marwan}{Rawald
  et~al.}{2017}]{Rawald2017}
Rawald T.,  Sips M.,   Marwan N.,  2017, \mn@doi [Computers and Geosciences]
  {10.1016/j.cageo.2016.11.016}, 104, 101

\bibitem[\protect\citeauthoryear{Robinson \& Thiel}{Robinson \&
  Thiel}{2009}]{Robinson2009}
Robinson G.,  Thiel M.,  2009, Chaos, 19

\bibitem[\protect\citeauthoryear{Ross, Latter  \& Tehranchi}{Ross
  et~al.}{2017}]{Ross2017}
Ross J.,  Latter H.~N.,   Tehranchi M.,  2017, \mn@doi [MNRAS]
  {10.1093/mnras/stx564}, 468, 2401

\bibitem[\protect\citeauthoryear{Ruelle}{Ruelle}{1978}]{Ruelle1978}
Ruelle D.,  1978, \mn@doi [Boletim da Sociedade Brasileira de Matem{\'{a}}tica]
  {10.1007/BF02584795}, 9

\bibitem[\protect\citeauthoryear{Runnoe, Brotherton  \& Shang}{Runnoe
  et~al.}{2012}]{Runnoe2012}
Runnoe J.~C.,  Brotherton M.~S.,   Shang Z.,  2012, \mn@doi [MNRAS]
  {10.1111/j.1365-2966.2012.20620.x}, 422, 478

\bibitem[\protect\citeauthoryear{Sauer, Yorke  \& Casdagli}{Sauer
  et~al.}{1991}]{Sauer1991}
Sauer T.,  Yorke J.~A.,   Casdagli M.,  1991, J. of Stat. Phys., 65

\bibitem[\protect\citeauthoryear{Scaringi et~al.,}{Scaringi
  et~al.}{2015}]{Scaringi2015}
Scaringi S.,  et~al., 2015, \mn@doi [Science Advances]
  {10.1126/sciadv.1500686}, 1

\bibitem[\protect\citeauthoryear{Schreiber \& Schmitz}{Schreiber \&
  Schmitz}{1996}]{Schreiber1996}
Schreiber T.,  Schmitz A.,  1996, PRL, 77, 635

\bibitem[\protect\citeauthoryear{Schreiber \& Schmitz}{Schreiber \&
  Schmitz}{2000}]{Schreiber2000}
Schreiber T.,  Schmitz A.,  2000, Physica D, 142

\bibitem[\protect\citeauthoryear{Shakura \& Sunyaev}{Shakura \&
  Sunyaev}{1973}]{Shakura1973}
Shakura N.,  Sunyaev R.~A.,  1973, A{\&}A, 24, 337

\bibitem[\protect\citeauthoryear{Small \& Judd}{Small \&
  Judd}{1998}]{Small1998}
Small M.,  Judd K.,  1998, Physica D, 120, 386

\bibitem[\protect\citeauthoryear{Small \& Judd}{Small \&
  Judd}{1999}]{Small1999}
Small M.,  Judd K.,  1999, \mn@doi [Phys. Rev. E] {10.1103/PhysRevE.59.1379},
  59, 1379

\bibitem[\protect\citeauthoryear{Small \& Tse}{Small \& Tse}{2002}]{Small2002}
Small M.,  Tse C.~K.,  2002, Physica D, 164, 187

\bibitem[\protect\citeauthoryear{Small \& Tse}{Small \& Tse}{2003}]{Small2003}
Small M.,  Tse C.~K.,  2003, \mn@doi [IEEE Transactions on Circuits and Systems
  I] {10.1109/TCSI.2003.811020}, 50, 663

\bibitem[\protect\citeauthoryear{Smirnov}{Smirnov}{1939}]{Smirnov1939}
Smirnov N.~V.,  1939, Byull. Mosk. Gos. Univ., i, 2

\bibitem[\protect\citeauthoryear{Smith et~al.,}{Smith et~al.}{2015}]{Smith2015}
Smith K.~L.,  et~al., 2015, \mn@doi [AJ] {10.1088/0004-6256/150/4/126}, 150,
  126

\bibitem[\protect\citeauthoryear{Smith, Mushotzky, Boyd, Malkan, Howell  \&
  Gelino}{Smith et~al.}{2018a}]{Smith2018a}
Smith K.~L.,  Mushotzky R.~F.,  Boyd P.~T.,  Malkan M.,  Howell S.~B.,   Gelino
  D.~M.,  2018a, \mn@doi [ApJ] {10.3847/1538-4357/aab88d}, 857, 141

\bibitem[\protect\citeauthoryear{Smith, Mushotzky, Boyd  \& Wagoner}{Smith
  et~al.}{2018b}]{Smith2018b}
Smith K.~L.,  Mushotzky R.~F.,  Boyd P.~T.,   Wagoner R.~V.,  2018b, \mn@doi
  [ApJ] {10.3847/2041-8213/aac88c}, 860, L10

\bibitem[\protect\citeauthoryear{Stella \& Vietri}{Stella \&
  Vietri}{1998}]{Stella1998}
Stella L.,  Vietri M.,  1998, ApJ, 492, 59

\bibitem[\protect\citeauthoryear{Stella, Vietri  \& Morsink}{Stella
  et~al.}{1999}]{Stella1999}
Stella L.,  Vietri M.,   Morsink S.~M.,  1999, ApJ, 524, 63

\bibitem[\protect\citeauthoryear{Stiele, Belloni, Kalemci  \& Motta}{Stiele
  et~al.}{2013}]{Stiele2013}
Stiele H.,  Belloni T.~M.,  Kalemci E.,   Motta S.,  2013, \mn@doi [MNRAS]
  {10.1093/mnras/sts548}, 429, 2655

\bibitem[\protect\citeauthoryear{Sukov{\'{a}}, Grzedzielski  \&
  Janiuk}{Sukov{\'{a}} et~al.}{2016}]{Sukova2016}
Sukov{\'{a}} P.,  Grzedzielski M.,   Janiuk A.,  2016, \mn@doi [A{\&}A]
  {10.1051/0004-6361/201526692}, 586

\bibitem[\protect\citeauthoryear{Tagger \& Pellat}{Tagger \&
  Pellat}{1999}]{Tagger1999}
Tagger M.,  Pellat R.,  1999, A{\&}A, 349, 1003

\bibitem[\protect\citeauthoryear{Takens}{Takens}{1981}]{Takens1981}
Takens F.,  1981, \mn@doi [Lecture Notes in Mathematics, Berlin Springer
  Verlag] {10.1007/BFb0091924}, 898, 366

\bibitem[\protect\citeauthoryear{{Terrell N. James}}{{Terrell N.
  James}}{1972}]{Terrell1972}
{Terrell N. James} J.,  1972, \mn@doi [ApJL] {10.1086/180944}, 174, L35

\bibitem[\protect\citeauthoryear{Theiler}{Theiler}{1986}]{Theiler1986}
Theiler J.,  1986, \mn@doi [Phys. Rev. A] {10.1103/PhysRevA.34.2427}, 34, 2427

\bibitem[\protect\citeauthoryear{Theiler \& Prichard}{Theiler \&
  Prichard}{1996}]{Theiler1996}
Theiler J.,  Prichard D.,  1996, Physica D, 94, 221

\bibitem[\protect\citeauthoryear{Theiler, Eubank, Longtin, Galdrikian  \&
  Farmer}{Theiler et~al.}{1992}]{Theiler1992}
Theiler J.,  Eubank S.,  Longtin A.,  Galdrikian B.,   Farmer J.~D.,  1992,
  Physica D, 58, 77

\bibitem[\protect\citeauthoryear{Thiel, Romano, Kurths, Meucci, Allaria  \&
  Arecchi}{Thiel et~al.}{2002}]{Thiel2002}
Thiel M.,  Romano M.~C.,  Kurths J.,  Meucci R.,  Allaria E.,   Arecchi F.~T.,
  2002, \mn@doi [Physica D] {10.1016/S0167-2789(02)00586-9}

\bibitem[\protect\citeauthoryear{Thiel, Romano  \& Kurths}{Thiel
  et~al.}{2003}]{Thiel2003}
Thiel M.,  Romano M.~C.,   Kurths J.,  2003, Izvestija VUZov—Applied
  Nonlinear Dynamics, 11, 20

\bibitem[\protect\citeauthoryear{Thiel, Romano, Read  \& Kurths}{Thiel
  et~al.}{2004}]{Thiel2004}
Thiel M.,  Romano M.~C.,  Read P.~L.,   Kurths J.,  2004, \mn@doi [Chaos]
  {10.1063/1.1667633}, 14, 234

\bibitem[\protect\citeauthoryear{Titarchuk \& Fiorito}{Titarchuk \&
  Fiorito}{2004}]{Titarchuk2004}
Titarchuk L.,  Fiorito R.,  2004, ApJ, 612, 988

\bibitem[\protect\citeauthoryear{Uttley \& McHardy}{Uttley \&
  McHardy}{2001}]{Uttley2001}
Uttley P.,  McHardy I.~M.,  2001, \mn@doi [MNRAS]
  {10.1046/j.1365-8711.2001.04496.x}, 323

\bibitem[\protect\citeauthoryear{Uttley, {M C Hardy}  \& Papadakis}{Uttley
  et~al.}{2002}]{Uttley2002}
Uttley P.,  {M C Hardy} I.~M.,   Papadakis I.~E.,  2002, MNRAS, 332, 231

\bibitem[\protect\citeauthoryear{Veledina, Poutanen  \& Ingram}{Veledina
  et~al.}{2013}]{Veledina2013}
Veledina A.,  Poutanen J.,   Ingram A.,  2013, \mn@doi [ApJ]
  {10.1088/0004-637X/778/2/165}, 778, 165

\bibitem[\protect\citeauthoryear{V{\'{e}}ron-Cetty \&
  V{\'{e}}ron}{V{\'{e}}ron-Cetty \& V{\'{e}}ron}{2003}]{VCV}
V{\'{e}}ron-Cetty M.~P.,  V{\'{e}}ron P.,  2003, \mn@doi [A{\&}A]
  {10.1051/0004-6361:20034225}, 412, 399

\bibitem[\protect\citeauthoryear{Voges, Atmanspacher  \& Scheingraber}{Voges
  et~al.}{1987}]{Voges1987}
Voges W.,  Atmanspacher H.,   Scheingraber H.,  1987, \mn@doi [J. of Chem.
  Info. and Modeling] {10.1017/CBO9781107415324.004}, 320, 794

\bibitem[\protect\citeauthoryear{Webber \& Zbilut}{Webber \&
  Zbilut}{1994}]{Webber1994}
Webber C.~L.,  Zbilut J.~P.,  1994, \mn@doi [J. of App. Physiology]
  {10.1152/jappl.1994.76.2.965}, 76, 965

\bibitem[\protect\citeauthoryear{Webber, Marwan, Facchini  \& Giuliani}{Webber
  et~al.}{2009}]{Webber2009}
Webber C.~L.,  Marwan N.,  Facchini A.,   Giuliani A.,  2009, \mn@doi [Phys.
  Lett. A] {10.1016/j.physleta.2009.08.052}, 373, 3753

\bibitem[\protect\citeauthoryear{Whitney}{Whitney}{1935}]{Whitney1935}
Whitney H.,  1935, \mn@doi [Proceedings of the NAS] {10.1073/pnas.21.7.462}

\bibitem[\protect\citeauthoryear{Wiita}{Wiita}{1996}]{Wiita1996}
Wiita P.~J.,  1996, in Miller H.~R.,  Webb J.~R.,   Noble J.~C.,  eds, Blazar
  Continuum Variability, ASP Conference Series.

\bibitem[\protect\citeauthoryear{Zbilut \& Marwan}{Zbilut \&
  Marwan}{2008}]{Zbilut2008}
Zbilut J.~P.,  Marwan N.,  2008, \mn@doi [Phys. Lett. A]
  {10.1016/j.physleta.2008.09.027}, 372, 6622

\bibitem[\protect\citeauthoryear{Zbilut \& Webber}{Zbilut \&
  Webber}{1992}]{Zbilut1992}
Zbilut J.,  Webber C.,  1992, \mn@doi [Phys. Lett. A]
  {10.1016/0375-9601(92)90426-M}, 171, 199

\bibitem[\protect\citeauthoryear{Zbilut, Zaldivar-Comenges  \& Strozzi}{Zbilut
  et~al.}{2002}]{Zbilut2002}
Zbilut J.~P.,  Zaldivar-Comenges J.-M.,   Strozzi F.,  2002, Phys. Lett. A,
  297, 173

\bibitem[\protect\citeauthoryear{Zou, Paz{\'{o}}, Romano, Thiel  \& Kurths}{Zou
  et~al.}{2007}]{Zou2007}
Zou Y.,  Paz{\'{o}} D.,  Romano M.~C.,  Thiel M.,   Kurths J.,  2007, \mn@doi
  [Phys. Rev. E] {10.1103/PhysRevE.76.016210}, 76

\makeatother
\end{thebibliography}



\appendix
\section{Recurrence Analysis: An Overview} \label{sec:recurrence}

The concept of recurrent behaviour in time series was first introduced by \citep{Poincare}, with visualisation of recurrences in the form of Poincar\'{e} plots, and with return maps \citep{Hilborn}. Recurrence Plots (RPs) were introduced by \cite{Eckmann1987} as a more general means to visualise the recurrences of trajectories embedded in phase space within dynamical systems. RPs provide qualitative information about the behaviour of the system of study, particularly indications of stochastic, periodic, or chaotic behaviour. Measures that quantify the structures present in RPs were introduced by \cite{Webber1994} and subsequently applied to various fields including Mathematics, Geology, and Physiology (\citealt{Gao2000}, \citealt{Marwan2002}, \citealt{Zbilut2002}, respectively) and others, where several of the quantitative measures, particularly those based on diagonal-line features, were mathematically equated to a variety of dynamical invariants underlying the observed time series. It is also possible to reconstruct a phase space from one-dimensional observations without loss of dynamical information \citep{Sauer1991}. Recurrences that appear in phase space also contain all the information about the dynamics of a system and constitute an alternative, and complete, description of a dynamical system \citep{Robinson2009}.

Recurrence analysis has recently been brought to astrophysics where RPs were utilised in the study of the stability of terrestrial planets \citep{Asghari2004}, in the distinction of chaotic and stochastic behaviour in the X-ray variability of microquasars \citep{Sukova2016}, and in the identification of non-linearity in the 14-year X-ray light curve of an XRB \citep{Phillipson2018}. For an extensive overview of the history of RPs, RQA, and their applications, see the seminal review by \cite{Marwan2007}. See also the appendices of \cite{Sukova2016} for a very similar and detailed approach to computing the entropy and determining embedding parameters. 

RPs are the graphical representation of the binary recurrence matrix, Eq.~\ref{eq:rp_mat1}, where a color represents each entry of the matrix (e.g., a black dot for unity and empty for zero). By convention, the axes increase in time rightwards and upwards. The RP is also symmetric about the main diagonal, called the `line of identity' (LOI). The recurrence matrix is computed after the time series is embedded in phase space. In the following sections, we describe phase space and approaches for determining the embedding parameters, followed by the quantification of structure that is seen in the recurrence plot.

\subsection{Phase Space} \label{subsubsec:PhaseSpace}

In order to compute the distances (often determined by the Euclidean metric) between two positions in a time series, which make up the entries of the recurrence matrix, we must first embed the time series in phase space. The most commonly employed embedding is a differential phase space, where the vectors in phase space represent successive derivates of an observable (e.g., position versus velocity). This type of embedding is also the most intuitive, because each component has an obvious relationship to the differential equation that describes the system in question, while also often having a physical interpretation. For example, the differential phase space embedding for a simple pendulum is the angular position versus angular velocity, which traces out a closed curve (e.g. a circle). 

For a damped or driven oscillator, the structure in differential phase space becomes more complicated. As we increase the damping and driving parameters, we might see ellipses in phase space that do not close upon themselves, resulting in recurrent but not periodic behaviour, and possibly entering into non-linear regimes with more complex trajectories. The relative structure of the nearly closed trajectories in phase space are unique to specific equations of motion and constitute a `topological' perspective of the dynamics. Topological information thus describes how a neighbourhood of points in phase space evolves in time, and how positions in phase space are globally organised, relative to each other \citep{Gilmore1998}. In fact, the way in which a neighbourhood of points is organised topologically is invariant under transformations --- topological information such as how various trajectories are organised relative to each other do not change through a variety of different types of embeddings (differential, or otherwise).

Given that we are dealing with a single observable, the flux, we must construct the phase space from the one-dimensional light curve. Furthermore, we do not know, or have an indication of, the dimension of the underlying attractor that generates the light curve and so a differential embedding is not possible. A commonly used method for reconstruction in this scenario is via the time delay method (\citealt{Whitney1935}, \citealt{Takens1981}), which reproduces the topological structure of the attractor from a single observable. The point here is that the differential information is invariant under transformations into higher-order spaces, even ones involving a single generic observable \citep{Sauer1991}. 

Following notation from \cite{Gilmore1998} (and \citealt{Phillipson2018}), for the scalar and discrete time series, $x(t)$, like a light curve, we construct vectors $y(t)$ with $n$ components. This involves creating an $n$ vector by the map
\begin{equation}
\begin{split}
	x(t) \rightarrow & y(t) = (y_1(t), y_2(t), \dots, y_n(t)) \\
				& y_k(t) = x(t-\tau_k), \,\,\, k=1,2,\dots,n,
\end{split}	
\label{eq:timedelay}
\end{equation}
where $\tau_k$ are called the time delays. The time delays are typically evenly spaced multiples of $the$ time delay: $\tau_k = (k-1)\tau$. 

Note that the embedding dimension using the time delay method does not carry physical units (much like the new vectors under principal component analysis do not carry the same literal meaning as the original data). We are no longer dealing with direct differentials of the original time series, even though the newly constructed vectors, $y(t)$, are constructed from flux values in the original light curve. The term ``embedding'' by definition means that the map from the space that contains the attractor into a reconstructed phase space is one-to-one and preserves differential information. In particular, \cite{Sauer1991} showed that an attractor with box-counting dimension $d$ can be reconstructed into a new space, $R^{m}$, via $m$ time-delayed versions of one generic observation, where $m \geq 2d+1$ is a requirement. For example, for an attractor that exists in a space with box-counting dimension of 1.4, we would require an embedding dimension of at least 4 in order to unambiguously reconstruct the topological information in a new space, where each component of the new embedded vector consists of values from the original time series spaced by the time delay. We also point out that the requirement on the dimension is an inequality, and so an even higher embedding dimension (e.g., 5 or 6) would also be appropriate and recover the same topological information -- indeed, many dynamical invariants are indifferent towards the embedding parameters \citep{Thiel2004}.

Recurrence plots, and other phase space based methods, reveal the dynamical information that generates the time series, but from a topological perspective. And since a time delay construction of phase space is an embedding, each of the $m$ components of the embedded system reflect the global organisation in phase space. This concept is readily evident for a simple pendulum embedded in a differential phase space: the points are organised in a circle, reflecting the periodic nature of the system. Recurrences probe this cyclical kind of global organisation. An embedding of a pendulum constructed from only the positional information will carry the same cyclical information of the attractor, even with an embedding dimension greater than 2. The usefulness of recurrence plots comes from this critical concept, as topological information is a powerful and direct discriminant of different dynamical systems.

There are a variety of approaches for determining the optimal embedding parameters. For determining the time delay, one can use the first minimum in the mutual information function (\citealt{Fraser1986}; \citealt{Pompe1993}). Mutual information operates similarly to an autocorrelation function, where the extrema provide us with information about correlations. However, mutual information also contains non-linear correlations, and is thus ideal for determining appropriate time delays when non-linearity might be a factor. Alternatively, one can use the first zero-crossing of the autocorrelation function -- the point is to ensure the values we use to construct a new $m$-dimensional embedded vector are not correlated to one another.

For choosing an appropriate embedding dimension, we seek to reconstruct the attractor such that no two trajectories cross each other, as self-intersections would violate the rule that there are unique solutions to the equations of motion. The false nearest-neighbours algorithm \citep{Kennel1992} is commonly used to determine the embedding dimension appropriate to avoid self-intersections. The false nearest-neighbours algorithm determines the minimal sufficient embedding dimension, $m$, such that neighbouring points in an embedded phase space represent neighbours in the fully reconstructed attractor. In practice, when we increase the embedding dimension, neighbouring points that then diverge from each other when you increase the dimension are called ``false neighbours.'' The algorithm consists of systematically increasing the embedding dimension until we reach a minimal number of false neighbours (theoretically when this value becomes zero; conventionally when less than 10 per cent). The resulting dimension is the minimum embedding dimension required to recover unambiguous topological information according to Takens' theorem \citep{Takens1981}. 

Using the mutual information function and the false nearest-neighbours algorithm from the non-linear dynamics software {\tt TISEAN}, we determine a minimum embedding dimension of $m=5$ is appropriate for both objects (resulting in less than 10 per cent false neighbours, though we could have used an even higher dimension), and a time delay of 27 days and 22 days for KIC 9650712 and Zw 229--015, respectively (note these time delays are comparable to the de-correlation times). Both these and other methods of determining the embedding parameters are reviewed by \cite{Marwan2007}. The Takens' approach for embedding a scalar time series via the time delay method is only one approach for embedding. Other embedding methods include performing singular value decomposition (\citealt{Sauer1991}, \citealt{Broomhead1986}), or independent component analysis \citep{Hyvarinen2001}, which similarly transform time series into other-dimensional spaces.

Once the embedding parameters are determined, one of the required parameters for generating RPs is the threshold, $\epsilon$, which as a rule of thumb should not exceed the maximum phase space diameter of the time series embedded in phase space \citep{Zbilut1992}, i.e., the threshold should not exceed the maximum size of the reconstructed space. An example of the recurrence plots of KIC 9650712 and Zw 229--015 for a threshold corresponding to a recurrence rate of 30 per cent is shown in Fig.~\ref{fig:rps_both}. 

In practice, given that the observed light curve of an astronomical system is a superposition of a real signal and some observational noise, a method for extracting an optimal threshold to produce a recurrence plot is to exploit the linear scaling relationship that exists between the threshold value and the corresponding recurrence rate. That is, if we produce a recurrence plot for every threshold that corresponds to a range of recurrence rates (called an `un-thresholded' recurrence plot, as in Fig.~\ref{fig:QPO_rp_closereturns} and Fig.~\ref{fig:Zw229_rp_closereturns}), there will be a range of values for which there exists a linear relationship between threshold and recurrence rate, when cast in logarithmic space. Any threshold that exists in the linear region of a $\log$-$\log$ plot of threshold value versus recurrence rate will equivalently probe the dynamics of the system \citep{Zbilut2002} without being dominated by noise (too small a threshold) or false-positive recurrences (too large a threshold). A range of thresholds is also required for the determination of dynamical invariants (see entropy calculation in Sec.~\ref{sec:K2_RQA}).

\subsection{Recurrence Quantification Analysis \& Dynamical Invariants} \label{subsec:RQA}

The structures in RPs have been classified via several measures of complexity known collectively as `recurrence quantification analysis', or RQA. These measures are based on the recurrence point density, or recurrence rate, defined as 
\begin{equation}
	RR(\epsilon) = \frac{1}{N^2}\sum_{i,j=1}^N \mathbf{R}_{i,j}(\epsilon), 
\end{equation}
and the diagonal and vertical line structures and their distributions present in the RP. Here $N$ is the length of the time series and, in the limit $N \rightarrow \infty$, $RR(\epsilon)$ is the probability that a state recurs to its $\epsilon$-neighbourhood in phase space (e.g., a time series that is a straight line, for example, would produce a 100 per cent recurrence rate).

RQA measures link the variety of large scale patterns to specific behaviours of a system. For example, a gaussian noise process produces a RP that is dominated by uniformly distributed random, isolated points. As a consequence, there are few uninterrupted lines throughout the RP and we therefore expect that a distribution based on diagonal line lengths, for example, would contain almost exclusively only short or unit-length lines. We can quantify the extent to which a recurrence plot is dominated by uncorrelated or weakly correlated behaviour (e.g., none or very short diagonal lines) through the ratio of recurrence points that form diagonal structures to all recurrence points, called the `determinism' (DET) of the system (or predictability):
\begin{equation}\label{eqn:det}
	DET = \frac{ \sum_{l=l_{min}}^{N} l P(l) }{ \sum_{l=1}^{N} l P(1) },
\end{equation}
where the lower limit $l_{min}$ is the minimum length line to consider, typically set to 2 \citep{Babaei2014}, and $P(l)$ is the histogram of diagonal line lengths (for a given threshold). 

Another important quantity relates to the lengths of the diagonal lines in the RP, whereby the longest line found, $L_{max}$, specifically relates to the exponential divergence of the phase space trajectory and, by consequence, the largest positive Lyapunov exponent \citep{Eckmann1987}, a quantity that characterises an attractor in studies of non-linear dynamics. In fact, the cumulative distribution of diagonal line lengths present in a RP is directly related to the correlation entropy (also known as the R\'{e}nyi entropy of second order, notated as $K_2$; \citealt{Grassberger1983}; \citealt{Thiel2003}), which is a measure of the complexity and predictability of the system and can be used to distinguish determinism from randomness \citep{Faure1998} and estimate various dimensions \citep{Grassberger1983}. A computation of the $K_2$ entropy is therefore useful for identifying non-linearity from linearity and stochasticity from determinism in a light curve and provides an estimate for the prediction horizon of the time series.

We have mentioned only a few of the RQA measures --- the recurrence rate, $RR$, the determinism, $DET$, and the longest diagonal line, $L_{max}$ --- that are most often used in the recurrence analysis literature (see the seminal review by \citealt{Marwan2007} for detailed definitions and applications). Diagonal lines trace recurrent behaviour in the time series whereby a state revisits itself at a later time. There are also measures based on the vertical lines and structures in a RP, which we discuss in Sec.~\ref{subsubsec: vertical} and Sec.~\ref{subsubsec: transitions} and apply to the KIC 9650712 and Zw 229--015 light curves. For purely periodic dynamics we expect the RQA measures based on vertical structures (such as the longest vertical line, $V_{max}$) to result in values of zero. In contrast, time series that instead contain regions with slowly changing, or unchanging, states will result in measurable vertical line structures in the RP. A distribution of the vertical line lengths can give indications about the length of time a particular state in a system will persist, or how long it will take to revisit a previous state. The vertical lines can provide information about fluctuations in the light curve, the rate at which they occur and how long they persist, in addition to state transitions (e.g. from chaos to periodicity or laminarity). For context, fluctuations are often related to turbulent mechanisms in the accretion disc.

Finally, the time dependence of RQA measures can be determined by computing these measures in small windows of the RP moving along the LOI. Windowed recurrence plots are particularly useful for detecting non-stationarity (i.e. fluctuations in the state parameters of the underlying system) and state transitions in the time series \citep{Marwan2002}. Such transitions could provide hints as to the dominating physical driving mechanisms underlying the light curve, or the times of their onset. For example, when $RR$ remains unchanged whereas $DET$ increases, a reorganisation of the recurrent points from isolated positions to an assembly of diagonal lines occurs, indicating a transition from noise to periodicity or quasi-periodicity; similarly, if the ratio of laminarty ($LAM$, ratio of recurrence points that form vertical structures to all recurrence points) to $DET$ can indicate transitions between dominant localised fluctuations to global, periodic or otherwise regular behaviour.

Other dynamical invariants, such as the $K_2$ entropy, based on either the diagonal or vertical line structures in a RP include generalised mutual information \citep{Thiel2003}, the correlation dimension, $D_2$ \citep{Grassberger1983}, and the point-wise dimension \citep{Gao1999}. The abstract concepts of dimension and entropy (relating to predictability and complexity of a time series) provide information about the family of differential equations that govern a particular system and give rise to the scalar time series of the observables. Translating to astronomy: entropy, dimension and other invariant measures provide evidence for the types of physical mechanisms that produce the different modes of variability in a light curve beyond merely the time-scales at which they occur.

\section{Numerical Method using Surrogate Data} \label{subsec:surrogates}

Recurrence analysis becomes a powerful distinguishing probe of various stochastic, deterministic and non-linear features in a time series when combined with the surrogate data method, introduced by \cite{Theiler1992} as an alternative, data-driven hypothesis test. In summary, a set of ``surrogate'' data are generated that resemble the original dataset. One then tests whether the original dataset is a member of the class of dynamical systems that generate the surrogates. The mode by which we generate the surrogate data represents a null hypothesis for the origin of the observed structures in the time series of interest. Once we have an ensemble of surrogate data sets, we can then look for additional structure that is present in the real data and not in the surrogates via a variety of statistical tests (in our case, using recurrence properties) which will either fail to rule out our null hypothesis as a good model for our data, or instead indicate that higher order modes or non-linear mechanisms are responsible for the features we observe. The number of surrogates that we generate dictates the level of confidence in our results \citep{Schreiber2000}.

To employ the surrogate data method, one must choose an appropriate null hypothesis for surrogate generation and an appropriate test statistic. The three algorithms introduced by \cite{Theiler1992} for surrogate generation still prominently used in the literature are based on Monte Carlo Fourier-based re-sampling techniques and generate surrogates that represent (i) independent and identically distributed noise (preserving the probability distribution), (ii) linearly filtered noise (preserving the power spectrum), and (iii) non-linear transformations of linearly filtered noise (preserving both the probability distribution and power spectrum). The methods of surrogate testing were initially introduced as a ``sanity" check for correlation dimension estimation (\citealt{Small1999}; \citealt{Small2002}; \citealt{Small2003}); estimations for correlation entropy and dimension remain a leading choice for the test statistic to distinguish noise and determinism with the surrogate data method (\citealt{Small1998}; \citealt{Sukova2016}; \citealt{Asghari2004}). The test statistic must be independent of the surrogate generation method (thus, computing the autocorrelation function or its derivatives would not be a good choice for the test statistic for surrogates generated via Fourier re-sampling techniques). 

In this study, we follow the same procedure introduced by \cite{Small2003}, where we estimate the correlation entropy ($K_2$ entropy) directly from the recurrence plot, a computation that is a function of viewing scale, hence $K_2(\epsilon)$. The result is an estimate for the entropy which is a curve, and not a single number, dependent on threshold. The test statistic for the comparison to the surrogate data generated will therefore be the deviation of the entropy computed by the data with respect to the surrogates, which can be interpreted graphically as a function of $\epsilon$. We note here that the full deterministic structure underlying the data can be seen only on length scales smaller than $\epsilon \sim e^{-h}$, where $h$ is the theoretical correlation entropy; above a critical length scale (i.e. for larger thresholds), the data and surrogates look equivalent, and if the critical $\epsilon$ is smaller than the noise level (e.g. approximately 5$\sigma$ of the photometric noise; \citealt{Thiel2002}), there is no way to distinguish signal from noise. Thus, our estimation of the $K_2$ entropy will be the most valid for the smallest $\epsilon$ values that rise above the noise \citep{Kantz}. 

The important distinction between the method of surrogate data and other statistical approaches is that it follows a ``constrained realisations'' approach \citep{Theiler1996} where we produce simulated time series generated by the original dataset itself, imposing the features of interest in addition to observational and dynamical noise on to the surrogates, rather than by fitting a statistical model to the data. Furthermore, if we compare our real data to multiple types of surrogate data representing an array of null hypotheses (e.g., \citealt{Small2003}; \citealt{Sukova2016}), we can narrow down the nature of the underlying dynamics producing the time series of our data. 

\section{Recovering $\tau_{corr}$ from the Diagonal Line Lengths}\label{subsubsec: tau_corr}

A characteristic time-scale can be recovered from the frequency of vertical lines in the recurrence plot (Fig.~\ref{fig:QPO_rp_closereturns} and Fig.~\ref{fig:Zw229_rp_closereturns}) that represents the ``recurrence period,'' or the average amount of time between successive states in a time series. It has been shown that the same time-scale can be recovered from the cumulative diagonal line histogram in (natural) logarithmic space (e.g., Fig.~\ref{fig:diagonal_histograms}) or from the autocorrelation function of the time series (\citealt{Thiel2003}, \citealt{Anishchenko2003}). For short line lengths, we observe a turnover of the slope in the $\log\,P_{\epsilon}^c(l)$ plot (Fig.~\ref{fig:diagonal_histograms}), used for the computation of the $K_2$ entropy. The cross-over time-scale separating the two scaling regions within the cumulative diagonal line distributions traces the de-correlation time-scale extracted from the autocorrelation function (ACF) \citep{Anishchenko2003}. The rate of the ACF decay in differential systems depends essentially on the structure of an attractor (containing the dynamics of the system) and on the influence of noise \citep{Anishchenko2003}. For certain non-hyperbolic attractor types, the autocorrelations decay exponentially. From the ACF, two time-scales can be distinguished, i.e. $\tau \leq \tau_{cor}$ and $\tau > \tau_{cor}$. In the first case the exponential decay is defined by fluctuations of the instantaneous amplitude, and in the second case it depends on the effective diffusion coefficient of the differential system (e.g., or for a Wiener process, the diffusion coefficient). 

In the cumulative distribution of diagonal line lengths, the steeper the slope in $\log\,P_{\epsilon}^c(l)$, the shorter the forecasting time (how far into the future we can reasonably predict the time series), which we would expect for the short-term fluctuations that occur for times $\tau \leq \tau_{cor}$, as it is much more difficult to predict behaviour for times less than the recurrence period. Meanwhile, for $\tau > \tau_{cor}$ we obtain time-scales that trace the long-term, deterministic dynamics (as also represented by mechanisms driving unstable periodic orbits as probed by the close returns plot) and result in the flatter slope for long line lengths in $\log\,P_{\epsilon}^c(l)$  corresponding to the R\'{e}nyi entropy of the second order, $K_2$ -- in this region, fluctuations in the light curve are no longer correlated to one another. Though the $K_2$ entropy is specifically defined in the limit of long line lengths, computing the slope for the line lengths in the region $\tau \leq \tau_{cor}$ can give an estimate for the forecasting time, or predictability, at smaller horizons (though it is not clear this region is related to any kind of dynamical invariant). The de-correlation time separating these two regions can also be related to the turnover time as described by an AR(1) (or DRW) process \citep{Kasliwal2015a}. Indeed, \cite{Kasliwal2017} found a de-correlation time-scale of $27.5$ days for Zw 229--015 (the transition in the cumulative distribution of diagonal line lengths for Zw 229--015 occurs at $22\pm3$ days).


\end{document}